\documentclass[11pt]{JHEP3}
\usepackage{epsf, cite, amsmath, amssymb}
\usepackage{epsfig}
\usepackage{slashed}

\newcommand{\p}{\partial}
\newcommand{\vx}{\vec{x}}

\newcommand{\Ot}{{\mathcal{O}}_\tau}
\newcommand{\Obt}{{\bar{\mathcal{O}}}_{\tau}}

\newcommand{\Db}{\bar{D}_{\Delta_{1}\Delta_{2}\Delta_{3}\Delta_{4}}}

\title{AdS/CFT for Four-Point Amplitudes involving Gravitino Exchange}
\author{Linda I. Uruchurtu \\Department of Applied Mathematics and Theoretical Physics
\\University of Cambridge \\Wilberforce Road, Cambridge CB3 0WA \\E-mail:
liu20@damtp.cam.ac.uk}

\abstract{ In this paper we compute the tree-level four-point
scattering amplitude of two dilatini and two axion-dilaton fields in
type IIB supergravity in $AdS_5\times S^{5}$.  A special feature of
this process is that there is an ``exotic'' channel in which there
are no single-particle poles.  Another novelty is that this process
involves the exchange of a bulk gravitino.  The amplitude is
interpreted in terms of ${\cal N}=4$ supersymmetric Yang--Mills
theory at large 't~Hooft coupling.  Properties of the Operator
Product Expansion are used to analyze the various contributions from
single- and double-trace operators in the weak and strongly coupled
regimes, and to determine the anomalous dimensions of semi-short
operators. The analysis is particularly clear in the exotic channel,
given the absence of BPS states.}

\preprint{DAMTP-2007-65 \\
          arXiv:0707.0424}
\keywords{AdS-CFT Correspondence, 1/$N$ Expansion, Strong Coupling Expansion}

\begin{document}

\section{Introduction}
According to the AdS/CFT correspondence \cite{Maldacena:1997re,Gubser:1998bc,Witten:1998qj},
the supergravity approximation to type II string theory in a $AdS_{5}\times S^{5}$
background corresponds to the strong 'tHooft
coupling limit of $SU(N)$ $\mathcal{N}=4$ supersymmetric Yang--Mills theory,
in the large $N$ limit.

The correspondence is easily checked for special protected processes that are independent
of the 't Hooft coupling, such as the two-point and three-point correlation functions
of BPS states, which are completely determined by superconformal symmetry.
The first non-trivial dependence on the coupling is observed in the correlation function
of four BPS operators, which provides non-trivial information
about the dynamics of the theory on the strongly coupled regime, via the correspondence (see
\cite{Heslop:2002hp} for references). Several examples have been worked out
explicitly \cite{Lee:1998bx,Arutyunov:2000py,D'Hoker:2000dm,Arutyunov:2002fh,Arutyunov:2000im,Arutyunov:2000ku,Bianchi:1998nk}.
These have given rise to various exact and
partial non-renormalization theorems for correlators in Yang-Mills
theory  (see \cite{D'Hoker:2002aw} for reviews), and have
shown that indeed there is a closed operator algebra in
$\mathcal{N}=4$ SYM theory, and the amplitudes can be given an operator product expansion (OPE)
interpretation \cite{D'Hoker:1999jp,Hoffmann:2000dx}.

In \cite{Arutyunov:2000ku} the correlation function of four superconformal
primaries (CPO's) in type IIB supergravity in $AdS_{5}\times S^{5}$ was evaluated
in the tree approximation and used to determine
the strongly coupled OPE of two operators in $\mathcal{N}=4$ SYM theory
in the large $N$ limit.  This expression was then compared with that
obtained in free field theory, and the results were used to gain evidence
for how the OPE expansion of two CPO's changes as the coupling varies. More explicitly,
the main idea in \cite{Arutyunov:2000ku} was to consider
the limit of the SYM correlator
\begin{equation}
\label{OPEexp} \langle {\cal{O}}_1 (\vx_1) {\cal{O}}_2 (\vx_2)
{\cal{O}}_3 (\vx_3) {\cal{O}}_4 (\vx_4) \rangle
\end{equation}
in which $\vx_1 \rightarrow \vx_2$ and $\vx_3 \rightarrow \vx_4$
simultaneously. This amounts to doing a double OPE expansion of the form
\begin{equation}
\label{OPEexp2} \langle {\cal{O}}_1 (\vx_1) {\cal{O}}_2 (\vx_2)
{\cal{O}}_3 (\vx_3) {\cal{O}}_4 (\vx_4) \rangle \sim \sum_{m,n}
\frac{C_{12m} C_{34n} \langle {\cal{O}}_m (\vx_2) {\cal{O}}_n
(\vx_4) \rangle}{\vert \vx_1 - \vx_2 \vert^{\Delta_1 +\Delta_2
-\Delta_m} \vert \vx_3 - \vx_4 \vert^{\Delta_3 +\Delta_4 -\Delta_n}}\cdots
\end{equation}
where we have only shown the leading singular terms explicitly and
$\Delta_i$ is the conformal dimension of ${\cal{O}}_i$. In this way,
from the computation of the four-point function in the supergravity approximation,
one can obtain the OPE of any two of the operators in strongly coupled
Yang-Mills at large $N$.
Comparison with the free field theory expression made it possible
to deduce some interesting features concerning the coupling constant
dependence of the anomalous dimensions of the intermediate single--trace
and double--trace operators that contribute in the $(\vx_{1}-\vx_{2})$
channel.

It follows that the OPE at strong coupling is very different
from the structure at weak coupling. On the free field theory there
are huge degeneracies of states with finite conformal dimensions,
but most of these are long (unprotected) states that develop
anomalous dimensions of $O((g_{YM}^2 N)^{1/4})$ at strong coupling
and completely decouple from the OPE. These are known to correspond,
on the gravity side of the correspondence,
to excited string states that disappear upon taking the supergravity
limit. There are also a number of states that pick up anomalous dimensions
but do not diverge in the supergravity limit $g_{YM}^2N \to \infty$.
Apart from the well-known  short operators, states of this type include
semi-short (double-trace) operators with dimensions that get corrections
of order $1/N^2$.

Similar OPE interpretations have been obtained for the case
of four superconformal scalar descendents (operators dual
to the type II dilaton-axion with $\Delta=4$) in
\cite{D'Hoker:1999jp,Hoffmann:2000dx}, but up to now there
has been no study of four-point correlation functions involving
fermionic operators in the superconformal multiplet (although
the exchange of a generic $1/2$-spin field
exchange was considered in \cite{Kawano:1999au}). Such correlators
will generically involve AdS diagrams with
an exchange of a massless gravitino in the bulk,
and this type of process has not been considered before, mainly because
of the absence of a viable expression for the massless propagator
\footnote{Although this propagator had already been considered in the literature
\cite{Grassi:2000dm}, technical issues arose when used in the computation of
a generic correlation function.} \cite{Basu:2006ti}.

Various techniques for computing four-point exchange diagrams were
developed by Freedman and D'Hoker in a series of papers
\cite{Freedman:1998bj,D'Hoker:1999pj}. Initially, evaluation of these diagrams
involved the integration over one AdS vertex of the bulk-to-bulk propagator
which was carried out by means of a tedious expansion and
resumation procedure. Later on, the same group developed a
new method \cite{D'Hoker:1999ni} which circumvented the explicit
evaluation of the integral, and did not required the knowledge of
the relevant bulk-to-bulk propagator, making the evaluation
a much simpler affair. The key point in this procedure
is the fact that the propagator couples to conserved currents and
satisfies an appropiate wave equation away from the source. In this paper,
we will extend this procedure to the case of a massless gravitino exchange
by studying the process
\begin{equation}
\langle \Ot(\vx_{1})\Lambda_i(\vx_{2}) \Obt(\vx_{3})\bar\Lambda^i(\vx_{4}) \rangle
\label{corrpaper}
\end{equation}
where $\Ot$ and $\Lambda_i$ are the
operators transforming under $\mathbf{1}$ and $\mathbf{\bar{4}}$ of
$SU(4)$, with conformal weights $\Delta=4$ and $\Delta=7/2$
respectively. These operators belong to the $1/2$-BPS current
multiplet, and are related to $\mathcal{O}_{2}$, the superconformal
primary, by supersymmetry transformations $\delta^{4}\mathcal{O}_{2}\sim\mathcal{O}_{\tau}$,
$\delta^{3}\mathcal{O}_{2}\sim\Lambda$ (with $\delta$ denoting a $Q$-transformation).

Although this process is related by supersymmetry to the four-point function
of CPO's, it has several novel features. In the supergravity
approximation it receives contributions from two exchange diagrams
-a graviton exchange in the $(\vx_{1}-\vx_{3})$ channel and a gravitino
exchange in the $(\vx_{1}-\vx_{4})$ channel-, together with a contact interaction. There
is no single--particle exchange in the $(\vx_{1}-\vx_{2})$ channel, which
follows from the $U(1)$ R-symmetry of supergravity \footnote{The $U(1)$ charge
in this channel is $7/2$, whereas the maximum charge of a single particle
is $2$ \cite{Schwarz:1983qr}}. We will refer to the $(\vx_{1}-\vx_{2})$ channel
as the ``exotic'' channel, in line with the older terminology for meson scattering.
Computation of this process will also allow us to study the OPE of fermionic operators,
and their behaviour under variation of the coupling. In particular, we will
focus on the OPE $\Ot(\vx_{1})\Lambda(\vx_{2})$, which is known to contain a
semi-short operator which is dual to a bound state in the gravity
theory. This semi-short operator may be of relevance for understanding
the truncation performed in \cite{Basu:2004dm}.

This paper is organized as follows. In Section 2 we summarize our results.
In Section 3 we derive the
effective action of type IIB supergravity, compactified on
$AdS_{5}\times S^{5}$, which is required for the computation and
compute the corresponding supergravity four-point
amplitude. We separately describe the contributions coming from
the graviton and gravitino exchanges, and the contact graph,
focusing on the generalization of the method \cite{D'Hoker:1999ni}
for the massless spin $3/2$ case. This result is in agreement with
the one obtained in \cite{Osborn:2007ho} using superconformal techniques
\footnote{We thank H. Osborn for discussing his results prior publication.}.
In Section 4 we will use free $\mathcal{N}=4$
SYM field theory to derive OPE's of relevance to the correlation function (\ref{corrpaper})
in various short-distance limits. This will be compared with the analogous limits
of the supergravity result, which corresponds to the strongly coupled gauge
theory. From this we will determine the fate of certain non-BPS states when the
theory becomes strongly coupled. The discussion and conclusions are
presented in Section 5. We also include two appendices that provide
technical details concerning $D$-functions, that enter the expression
for the amplitude.
\section{Summary of the supergravity amplitude}
The detailed evaluation of the three diagrams that
contribute to the amplitude will be given in section 3. For clarity
we will now summarize the results. The complete contribution
to the four-point function in the supergravity approximation, is given by
\begin{eqnarray}
\langle \Ot(\vx_{1})\Lambda(\vx_{2}) \Obt(\vx_{3})\bar\Lambda(\vx_{4}) \rangle&=&\frac{\slashed{\vx}_{24}}{|\vx_{13}|^8|\vx_{24}|^8}
\nonumber \\
&+&\frac{1}{3N^2}\frac{1}{|\vx_{13}|^8|\vx_{24}|^8}
\left[I_{graviton}+I_{gravitino}+I_{quartic}\right]
\end{eqnarray}
where the first term is the disconnected contribution, and the term of order $1/N^2$ gives the connected contribution. Here
\begin{eqnarray}
I_{graviton}&=& \Big[(u+v)\left(6\bar{D}_{2525}+8\bar{D}_{3535}
+5\bar{D}_{4545}\right)-2\bar{D}_{1414}-8\bar{D}_{2424}-8\bar{D}_{3434}
\nonumber \\
&+&\left. 11\bar{D}_{4444}-\frac{5}{2}\bar{D}_{5454}
-\frac{3}{2}(\bar{D}_{3526}-\bar{D}_{2536})-\frac{5}{2}(\bar{D}_{4536}-\bar{D}_{3546})\right]\slashed{\vx_{24}}
\nonumber \\
&+&\left[3(\bar{D}_{2536}-\bar{D}_{3526})+5(\bar{D}_{3546}-\bar{D}_{4536})\right]
\frac{\slashed{\vx_{23}}\slashed{\tilde{x}_{31}}\slashed{\vx_{14}}}{|\vx_{13}|^2}
\nonumber \\
I_{gravitino}&=& \left[
4\bar{D}_{2442}+6\bar{D}_{3443}-16\bar{D}_{4444}-6\bar{D}_{3452}-8\bar{D}_{4453}+10\bar{D}_{5454}\right]\slashed{\vx_{24}}
\nonumber \\
&+&\left[-6\bar{D}_{2552}
-12\bar{D}_{3553}-10\bar{D}_{4554}+2\bar{D}_{2451}+6\bar{D}_{3452}+6\bar{D}_{4453}\right]
\frac{\slashed{\vx_{23}}\slashed{\tilde{x}_{31}}\slashed{\vx_{14}}}{|\vx_{13}|^2}
\nonumber \\
I_{quartic}&=&[15\bar{D}_{5445}-30\bar{D}_{4444}]\slashed{\vec{x}_{24}}-
15\bar{D}_{5454}\frac{\slashed{\vec{x}_{23}}\slashed{\vec{x}_{31}}\slashed{\vec{x}_{14}}}{|\vec{x}_{13}|^2}
\end{eqnarray}
The $\bar{D}$-functions,
$\bar{D}_{\Delta_{1}\Delta_{2}\Delta_{3}\Delta_{4}}(u,v)$, are
standard four-point contact diagrams in AdS, involving the contact
interaction of four scalars of conformal dimensions $\Delta_{i}$
which depend on the conformal ratios
\begin{equation}
u=\frac{|\vx_{12}|^2|\vx_{34}|^2}{|\vx_{13}|^2|\vx_{24}|^2} \quad
v=\frac{|\vx_{14}|^2|\vx_{23}|^2}{|\vx_{13}|^2|\vx_{24}|^2}
\end{equation}
In section 4 we will analyze this amplitude in various short distance limits,
in order to determine properties of its intermediate states.
\section{Supergravity Four-Point Function}
The precise statement of the AdS/CFT correspondence that allows us to
compute correlation functions of Yang-Mills operators from the
supergravity AdS theory, was given by Witten in
\cite{Witten:1998qj}. In a schematic form, one has
\begin{equation}
\mathrm{exp}\left\{-S_{IIB}[\phi]\right\}=\left\langle
\mathrm{exp}\left(\int_{\p(AdS)}\mathcal{O}\phi_{0}\right)\right\rangle_{YM}
\label{adsrecipe}
\end{equation}
where $\phi$ denotes the solution of the supergravity equations of motion, and the
value of $\phi$ at the boundary of AdS, is identified with the source $\phi_{0}$
that couples to a conformal field $\mathcal{O}$.

We will focus on operators that belong to the well-known $1/2$ BPS
current multiplet \cite{Bergshoeff:1980is}, which are dual to fields
on the gravity supermultiplet of type IIB sugra on $AdS_{5}\times
S^{5}$. Some four-point function of these operators have been
computed before, namely, those involving the top operators of the
multiplet, or chiral primaries \cite{Arutyunov:2000py}, and the
bottom operators of the multiplet, which transform as singlets of
$SU(4)$ \cite{D'Hoker:1999pj}. In the present case, we will explore
the correlation function $\langle \Ot(\vx_{1})
\Lambda_i(\vx_{2}) \Obt(\vx_{3}) \bar\Lambda^i(\vx_{4}) \rangle$,
where $\Ot$ and $\Lambda_i$ are the operators transforming under
$\mathbf{1}$ and $\mathbf{\bar{4}}$ of $SU(4)$, and conformal
weights $\Delta=4$ and $\Delta=7/2$ respectively. Before proceeding
with the calculation, let us first analyze the supergravity fields
to which they are dual.

The operators $\Ot$ and $\Obt$ enter the Yang-Mills action as
\begin{equation}
S =\frac{i}{4\tau_2}\int d^4 x \Big[\tau \Ot -\bar\tau \Obt \Big]
\end{equation}
where $\tau=\frac{\theta_{YM}}{4\pi}+i\frac{4\pi}{g_{YM}^2}=\tau_{1}+i\tau_{2}$
is the complex coupling constant.
Hence one could naively think $\Ot$ is sourced by $\tau/\tau_2$ in supergravity,
where $\tau$ is the axion-dilaton complex scalar $\tau=C_{0}+ie^{\phi}$, but this
is not true as it does not transform as a modular form under the
the $SL(2,Z)$ symmetry. The source is in fact
$i (\delta \tau)/{\sqrt 2}\tau_2$ which is a modular form of weight $(-1,1)$. So
at the linearized level, $\Ot$ is sourced by $i (\delta \tau)/{\sqrt 2}
\tau_2^0 $ where
$\tau_2^0$ stands for the background value of $\tau_2$. In this way, the relevant field
for our computations is identified to be
\begin{equation}
P_{\mu} = \frac{i}{2}\frac{\p_{\mu} \tau}{\tau_2}
\end{equation}
The spin $1/2$ operators, $\Lambda_{i}$ and $\bar{\Lambda}^i$, are
easier to identify, and from the quantum numbers, one can see they
are sourced by the massless dilatino of the supergravity theory.
Having identified the bulk fields, one can start by finding the
relevant couplings from the IIB supergravity equations of motion
\cite{Schwarz:1983qr,Howe:1983sr}, and its Kaluza-Klein reduction to
$d=5$ \cite{PhysRevD.32.389}.
\subsection{Results of the Reduction}
We first write down the relevant terms in the $d=10$ type IIB
supergravity action, and then obtain the $d=5$ action. In the
absence of a simple action of this chiral supergravity that
implements the self--duality of the five--form field strength at the
level of the action, we consider the covariant equations of motion
which give us the relevant terms in the action. We focus on the
fermionic terms as the bosonic terms have been written down before
\cite{Arutyunov:1998hf}, and use a hat to denote ten dimensional fields.
Considering the linearized equations of
motion for the dilatino $\hat{\Lambda}$ and the gravitino $\psi_M$
given by (see (4.6) and (4.12) of \cite{Schwarz:1983qr}).
\begin{eqnarray}
&&\Gamma^M D_M \hat\Lambda -i\frac{\kappa_{10} g_s}{240} \Gamma^{M_1 \ldots M_5}
\hat\Lambda F_{M_1 \ldots M_5}=0  \nonumber \\
&&\Gamma^{MNP} D_N \psi_P -\frac{1}{4\tau_2}
(\p_N \tau) \Gamma^N \Gamma^M \hat\Lambda^* +i\frac{\kappa_{10} g_s}{480} \Gamma^{MNP}
\Gamma^{PQRST} \Gamma_N \psi_P F_{PQRST}=0
\end{eqnarray}
we get the action
\begin{eqnarray}
\label{tendaction}
S = \int d^{10} x {\sqrt{-g'}}&
\Big[& 2\hat{P}^{M}\hat{P}^{*}_{M}+\bar{\hat\Lambda} \Gamma^M D_M \hat\Lambda +{\bar\psi}_M \Gamma^{MNP} D_N
\psi_P
\nonumber \\
&-&i \frac{\kappa_{10} g_s}{240} \bar{\hat\Lambda} \Gamma^{M_1 \ldots M_5} \hat\Lambda
F_{M_1 \ldots M_5}
+\frac{i}{2} \hat{P}_{N} {\bar\psi}_M \Gamma^N \Gamma^M \hat\Lambda^* \nonumber \\
&+&\frac{i}{2} \hat{P}^{*}_N  \bar{\hat\Lambda}^* \Gamma^M \Gamma^N \psi_M
+i\frac{\kappa_{10} g_s}{480} {\bar\psi}_M \Gamma^{MNP}
\Gamma^{PQRST} \Gamma_N \psi_P F_{PQRST} \Big].
\end{eqnarray}
We consider the fields in the $SL(2,R)/U(1)$ formulation, in which
the axion-dilaton complex scalar $\hat{\tau}$ is contained in the $SL(2,R)$
singlet, $\hat{P}_{M}$, so in the $\varphi=0$ gauge \cite{Green:1998by}
takes the form
\begin{equation}
\hat{P}_{M}=\frac{i}{2}\frac{\p_{M}\hat{\tau}}{\hat{\tau_{2}}}
\end{equation}
The covariant derivative $D_{M}$ appearing in the action contains not only the spin-connection,
but also the $U(1)$ connection $Q_{M}$, so again in the $\varphi=0$ gauge,
the term appearing in the derivative
has the form $-q\frac{\p_{M}\hat{\tau}}{2\hat{\tau}}$, where $q$ is the $U(1)$ charge.

One can then obtain the relevant couplings from the Kaluza-Klein
reduction of (\ref{tendaction}) on $AdS_{5}\times S^{5}$. The
quadratic terms have been computed in \cite{Arutyunov:1998hf}, and
are given by
\begin{eqnarray}
S_{\rm{quad}}=\frac{1}{2\kappa_5^2} \int_{AdS_5} d^5 x {\sqrt{-g}} && \left[ R
-\frac{\vert \p_\mu \tau \vert^2}{2\tau_2^2} +\bar{\Lambda}_i \left( \Gamma^\mu D_\mu
+\frac{3}{2} \right) \Lambda^i \right.
\nonumber \\
&+&\left.{\bar\psi}_\mu^i
\left( \Gamma^{\mu\nu\rho} D_\nu \psi_{\rho i}
+ \frac{3}{2} \Gamma^{\mu\nu} \psi_{\nu i} \right)\right]
\label{Squad}
\end{eqnarray}
which yields a dilatino and a gravitino of mass $-3/2$ \footnote{Here we have set the radius of $AdS$ to one and
we follow the conventions specified in \cite{Arutyunov:1998hf}}.
Note that both masses satisfy the relation
\begin{equation}
|m|=\Delta-2
\end{equation}
since $\Delta=7/2$ for both dual operators in the gauge theory. We have also rescaled the fermion
fields so that the full action has an overall factor of $1/2\kappa_{5}^2$.

The cubic terms can also be obtained from reduction of the ten-dimensional action. These are given by
\begin{eqnarray}
S_{\rm{cubic}}
= -\frac{1}{2\kappa_{5}^{2}}\int_{AdS_5} d^5 x \sqrt{-g} &\Big[&\frac{1}{4\tau_2} (\p_\mu \tau) {\bar\psi}_\nu^i \Gamma^\mu
\Gamma^\nu \Lambda^*_i -\frac{1}{4\tau_2} (\p_\nu \bar\tau) \bar{\Lambda}^{*i} \Gamma^\mu
\Gamma^\nu \psi_{\mu i} \nonumber \\
&+&\frac{3}{8}i\bar{\Lambda}^{i}\Gamma^{\mu}\Lambda_{i}\frac{(\p_{\mu}\tau)}{\tau_{2}}
+\frac{3}{8}i\bar{\Lambda}^{i}\Gamma^{\mu}\Lambda_{i}\frac{(\p_{\mu}\bar{\tau})}{\tau_{2}} \Big]
\label{Scubic}
\end{eqnarray}
Here the terms in the second line are obtained from the $Q_{M}$ term
in the ten-dimensional covariant derivative, when acting on the
dilatino. The free theory fermionic actions in (\ref{Squad}) need to
be supplemented with certain boundary terms, as they vanish
on-shell. These terms are defined on closed four dimensional
submanifolds in $AdS_{5}$ in the limit when the submanifold
approaches the boundary. The role of these boundary terms for spin
1/2 fermions has been studied in
\cite{Henningson:1998cd,Muck:1998iz,Ghezelbash:1998pf,Arutyunov:1998ve}, while for
Rarita--Schwinger fields it has been analyzed in
\cite{Corley:1998qg,Volovich:1998tj,Koshelev:1998tu}. Although these terms will not
affect our calculation in any way, they are included for completeness.
These are
\begin{equation}
\label{Sbound}
S_{\rm{bound}}
= \frac{1}{2\kappa_{5}^{2}} \int_{\p AdS_5} d^4 \zeta \sqrt{-\bar{g}} \frac{1}{2}\left( \bar{\Lambda}_i
\Lambda^i - {\bar\psi}^i_{m'} {\bar{g}}^{m'n'} \psi_{in'} \right)
\end{equation}
where ${\bar{g}}_{m'n'}$ is the induced metric on the boundary.

We are now ready to compute the four-point function. As usual, we will work with the Euclidean version of
$AdS_{5}$, so that the action is given by
\begin{eqnarray}
\label{totact}
S &=& \frac{1}{2\kappa_5^2} \int_{EAdS_5} d^5 x {\sqrt{g}} \Big[
 -R +\frac{\vert \p_\mu \tau \vert^2}{2\tau_2^2}
+\bar{\Lambda}_i \left( \Gamma^\mu D_\mu
+\frac{3}{2} \right) \Lambda^i \nonumber \\
&+&{\bar\psi}_\mu^i \left( \Gamma^{\mu\nu\rho} D_\nu \psi_{\rho i} +
\frac{3}{2} \Gamma^{\mu\nu} \psi_{\nu i} \right) -\frac{1}{4\tau_2}
\Big( (\p_\mu \tau) {\bar\psi}_\nu^i \Gamma^\mu \Gamma^\nu
\Lambda^*_i -(\p_\nu \bar\tau) \bar{\Lambda}^{*i} \Gamma^\mu
\Gamma^\nu \psi_{\mu i} \Big) \Big. \nonumber \\
&-&\Big.\frac{3i}{8\tau_{2}}\left(
\bar{\Lambda}_{i}\Gamma^{\mu}(\partial_{\mu}\tau)\Lambda^{i}
+\bar{\Lambda}_{i}\Gamma^{\mu}(\partial_{\mu}\bar{\tau})\Lambda^{i}\right)\Big]
\nonumber \\
&-&\frac{1}{2} \int_{\p EAdS_5}  d^4 \zeta \sqrt{\bar{g}} \left(
\bar{\Lambda}_i \Lambda^i - {\bar\psi}^i_{m'}
{\bar{g}}^{m'n'} \psi_{in'} \right).
\end{eqnarray}
The five dimensional axion and the dilaton are obtained in the obvious way
from those in ten dimensions, while the graviton is obtained from the ten dimensional one using the relation
$h_{\mu\nu} = h'_{\mu\nu} +\frac{1}{3} (g_0)_{\mu\nu} h'^\alpha_\alpha$, where $g_0$ is
the $AdS_5$ metric and $h'^\alpha_\alpha$ is the trace of the graviton on
$S^5$. Here
\begin{equation}
\frac{1}{2\kappa_5^2} =\frac{{\rm Vol}(S^5)}{2 g_s^2 \kappa_{10}^2}
=\frac{\pi^3 R^5}{g_s^2 (2\pi)^7 \alpha'^4} =\frac{N^2}{8\pi^2 R^3}
\end{equation}
\begin{figure}
 \centering
 \includegraphics[width=0.9\textwidth]{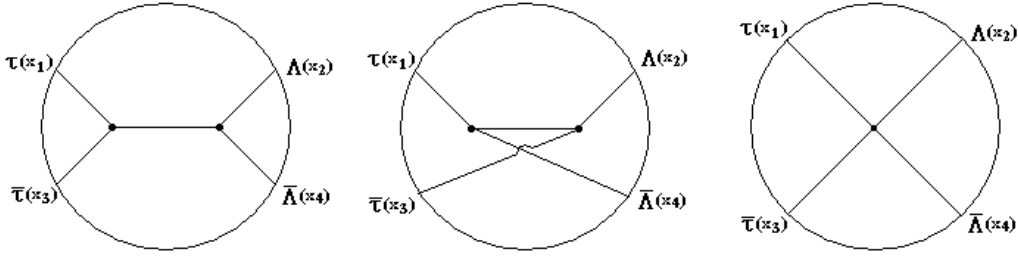}
 \caption{Witten diagrams contribuing to the process.}
 \label{figurediag}
\end{figure}
We can now see directly that the correlator will have one graviton
exchange and one gravitino exchange in the bulk from the $t$ and $u$
channels respectively. There is also a contact diagram, coming from
the quartic coupling that is contined in the
$-\frac{3}{4}\bar{\hat{\Lambda}}_{i}\Gamma^{M}Q_{M}\hat{\Lambda}^{i}$
term in the Dirac action for $\hat{\Lambda}^{i}$. Figure 1 depicts the
relevant diagrams contributing to the process. One should notice that
this is all in agreement with the $U(1)$ symmetry of supergravity.
The computation of these diagrams will be carried out in the following
sections.
\subsection{Graviton Exchange}
As in most previous work, we work on Euclidean $AdS_{5}$
defined as the upper half space in
$z^\mu \in \mathbf{R}^{5}$, with $z^{0} > 0$, with the metric given by
\begin{equation}
ds^{2}=\sum_{\mu,\nu=0}^{d} g_{\mu\nu}dz^{\mu}dz^{\nu}
=\frac{1}{z_{0}^{2}}(dz_{0}^{2}+\sum^{d}_{i=1}dz_{i}^{2})
\end{equation}
The coordinates $z_\mu$ will be raised and lowered with the
flat space metric unless otherwise
mentioned. We shall choose the vielbein to be given by
\begin{equation}
e_{\mu}^{a}=\frac{1}{z_{0}}\delta_{\mu}^{a}
\end{equation}
so that the spin and Levi-Civita connections are given by
\begin{eqnarray}
w_{\mu}^{ab}&=&\frac{1}{z_0}
(\delta^{a}_{0}\delta^{b}_{\mu}-\delta^{b}_{0}\delta^{a}_{\mu}) \nonumber \\
\Gamma^{\rho}_{\mu\nu}&=&\frac{1}{z_{0}}(\delta^{\rho}_{0}\delta_{\mu\nu}
-\delta_{\mu 0}\delta^{\rho}_{\nu}-\delta_{\nu 0}\delta^{\rho}_{\mu})
\end{eqnarray}
where $a,b$ are tangent indices. The Dirac matrices in curved space
will be related to those in the tangent space by $\Gamma^\mu
=e^\mu_a \gamma^a$, so that $\{ \gamma^a, \gamma^b \}
=2\delta^{ab}$. We now proceed to work out the amplitudes
explicitly.

The supergravity calculation that we wish to address in this section,
is the process involving an exchange of a massless graviton between two
scalars and two spin $1/2$ fermions. In the setting of type IIB compactified
on $AdS_{5}\times S^{5}$, one can identify the scalar as the
axion-dilaton field and the fermion with the dilatino.

Now any supergravity operator $\hat{\cal{O}} (z)$ in the bulk can be
expressed in terms of its boundary value $J_{{\cal{O}}} (\vec{x})$
which acts as the source for the composite operator ${\cal{O}}
(\vec{x})$ in YM, as was indicated in \cite{Witten:1998qj}. The
explicit relation is given by
\begin{equation}
\hat{\cal{O}} (z) = \int d^d \vec{x} K_{\Delta}(z, \vec{x})
J_{{\cal{O}}} (\vec{x}),
\end{equation}
where $K_{\Delta}(z, \vec{x})$ is the expression for the
bulk--to--boundary propagator. For calculating the amplitude, the
various expressions that are relevant are
\begin{eqnarray}
i \frac{\delta\tau}{\sqrt{2}\tau_{2}}(z) &=& \int d^4
\vec{x} ~K_4 (z, \vec{x})
J_{\mathcal{O}_{\tau}} (\vec{x}) \nonumber \\
\Lambda_i (z) &=&\int d^4 \vec{x} ~K_{7/2} (z,\vec{x})
\frac{(1+\gamma_0)}{2} J_{i \Lambda} (\vec{x}) \nonumber \\
\bar{\Lambda}^i (z) &=&\int d^4 \vec{x} ~{\bar{J}}^i_{\Lambda}
(\vec{x}) \frac{(1-\gamma_0)}{2} K_{7/2} (z,\vec{x})
\label{bbprops}
\end{eqnarray}
Here $K_4 (z,\vec{x})$ is the normalized bulk-to-boundary scalar
propagator, which is given in general by
\begin{equation}
K_{\Delta} (z,\vec{x})= C_{\Delta} {\widetilde{K}}_\Delta (z
,\vec{x})=\frac{\Gamma(\Delta)}{\pi^{d/2}\Gamma(\Delta-\frac{d}{2})}\frac{z_{0}^{\Delta}}{(z-\vec{x})^{2\Delta}},
\end{equation}
and $K_{7/2} (z,\vec{x})$ is the spin 1/2 fermion propagator
\begin{equation}
K_{7/2} (z,\vec{x}) = K_4 (z,\vec{x}) U (z -\vec{x})= K_4
(z,\vec{x}) \frac{(\slashed{z} -\slashed{\vec{x}})}{\sqrt z_0}
\end{equation}

The starting point is the effective action (\ref{totact}). One needs to represent the
solution to the equations of motion as $\phi=\phi^{0}+\delta\phi$ for a generic field,
where $\phi^{0}$ is the solution of the linearized equation of motion with fixed
boundary conditions, and $\delta\phi$ represents the physical field on the bulk. In
order to proceed, it is useful to introduce the following conserved currents
\begin{equation}
T^{\mu\nu}=\frac{1}{2\tau_{2}^{2}}\left(D^{(\mu}\tau D^{\nu)}\bar{\tau}
-\frac{1}{2}g^{\mu\nu} D^{\rho}\tau D_{\rho}\bar{\tau}\right) \qquad
\tilde{T}^{\mu\nu}=\bar{\Lambda}\Gamma^{(\mu}\overleftrightarrow{D}^{\nu)} \bar{\Lambda}
\end{equation}
One can write the graviton in terms of its Green function as
\begin{equation}
\delta g_{\mu\nu}(w)=\int [dz]G_{\mu\nu\mu'\nu'}(w,z)T^{\mu\nu}(z)
\end{equation}
so using this expression, we find the on-shell value of the action that is relevant
for this diagram
\begin{equation}
I_{graviton}=\frac{1}{4}\int [dz][dw] \tilde{T}^{\mu\nu}(w)G_{\mu\nu\mu'\nu'}(w,z)T^{\mu'\nu'}(z)
\end{equation}
The next step is to express the fields in terms of their boundary values (Eq. (\ref{bbprops})).
In this way the contribution to the amplitude is given by
\begin{equation}
I_{graviton}=\frac{1}{4}\int [dw]
\tilde{T}^{\mu\nu}(w,\vx_{2},\vx_{4})A_{\mu\nu}(w,\vx_{1},\vx_{3})
\label{gravitonex2}
\end{equation}
\begin{equation}
A_{\mu\nu}(w,\vx_{1},\vx_{3})=\int [dz]
G_{\mu\nu\mu'\nu'}(w,z)T^{\mu'\nu'}(w,\vx_{1},\vx_{3})
\end{equation}
where the vertex factors, are given by
\begin{eqnarray}
T^{\mu\nu}(z,\vx_{1},\vx_{3})&=&D^{\mu}K_{4}(w,\vx_{1})D^{\nu}K_{4}(w,\vx_{3})
-\frac{1}{2}g^{\mu\nu}\p_{\rho}K_{4}(w,\vx_{1})D^{\rho}K_{4}(z,\vx_{3})
\nonumber \\
\tilde{T}^{\mu\nu}(w,\vx_{2},\vx_{4})&=&\frac{1}{2}\mathcal{P}_{+}K_{7/2}(w,\vx_{2})(\Gamma^{\mu}\overleftrightarrow{D}^{\nu}
+\Gamma^{\nu}\overleftrightarrow{D}^{\mu})K_{7/2}(w,\vx_{4})\mathcal{P}_{-}
\end{eqnarray}
We will use the methods developed in \cite{D'Hoker:1999ni} to
compute this diagram, which take advantage of the fact that the
vertices are covariantly conserved. By inversion, one can see that
the $z$-integral can be expressed as
\begin{equation}
A_{\mu\nu}(w,\vx_{1},\vx_{3})=C_{4}^2|\vx_{31}|^{-8}\frac{J_{\mu\lambda}(w)}{w^2}
\frac{J_{\nu\rho}(w)}{w^2}I_{\lambda\rho}(w'-\vx_{31}')
\end{equation}
where $J_{\mu\nu}(w)=\delta_{\mu\nu}-2\frac{w_{\mu}w_{\nu}}{w^2}$ is
the conformal jacobian and the primes denote inverted coordinates,
$z_{\mu}'=\frac{z_{\mu}}{z^2}$. The key idea for
performing this integral, is to make an ansatz for $I_{\mu\nu}(w)$
that respects conformal invariance, and use the Green's function for
the graviton propagator. This was carried out in
\cite{D'Hoker:1999ni}, and we reproduce the result here
\begin{equation}
I_{\mu\nu}(w)=-\frac{2}{3}(t+t^2+t^3)\left(
\frac{\delta_{0\mu}\delta_{0\nu}}{w_{0}^2}-\frac{1}{3}g_{\mu\nu}
\right)
\end{equation}
where $t=(w_{0}/w)^2$. Substituting this expression back into
(\ref{gravitonex2}), and inverting the other vertex,
$\tilde{T}^{\mu\nu}(w,\vx_{21},\vx_{41})$, we see that there are
two contributions to the amplitude, one proportional to $\tilde{T}_{00}$ and
one proportional to ${\tilde{T}^{\mu}}_{\mu}$. The first contribution reads
\begin{eqnarray}
I^{(1)}_{graviton}=-\frac{1}{6}\frac{C_{4}^2}{|\vx_{21}||^{6}\vx_{31}|^8|\vx_{41}|^6}
&&\int [dw]w_{0}(t+t^2+t^3)\left\{\mathcal{P}_{+}K_{4}(w,\vx_{21}')\slashed{\vx_{21}'}U(w-\vx_{21}') \right.\nonumber \\
&\times& \left.\left(\gamma_{0}\overleftrightarrow{\p_{0}}+\overleftrightarrow{\p_{0}}\gamma_{0}\right)
U(w-\vx_{41}')\slashed{\vx_{41}'}K_{4}(w,\vx_{41}')\mathcal{P}_{-}\right\}
\end{eqnarray}
We can simplify this expression by using the following identities
\begin{eqnarray}
\p_{0}{U(w-\vx')K_{4}(w,\vx')}&=&\frac{7}{2w_{0}}U(w-\vx')K_{4}(w,\vx')-4U(w-\vx')K_{5}(w,\vx')
+\frac{\gamma_{0}}{\sqrt{w_{0}}}K_{4}(w,\vx')
\nonumber \\
w_{0}\p_{0}\tilde{K}_{4}(w,\vx')&=&4\tilde{K}_{4}(w,\vx')-8w_{0}\tilde{K}_{5}(w,\vx')
\end{eqnarray}
Hence we get
\begin{eqnarray}
I^{(1)}_{graviton}&=&-\frac{1}{6}\frac{C_{4}^4}{|\vx_{21}|^{6}\vx_{31}|^8|\vx_{41}|^6}\slashed{\vx_{21}'}\int
[dw](t+t^2+t^3)\left\{
-\tilde{K}_{4}(w,\vx_{21}')\tilde{K}_{4}(w,\vx_{41}')\slashed{\vx_{24}'} \right.\nonumber \\
&+&
\left[\tilde{K}_{4}(w,\vx_{21}')w_{0}\p_{0}\tilde{K}_{4}(w,\vx_{41}')
-\vx_{21}'\leftrightarrow
\vx_{41}'\right]\times
\nonumber \\
&&\left.\left[(\slashed{\vec{w}}-\slashed{\vx_{21}'})+
(\slashed{\vec{w}}-\slashed{\vx_{41}'}) \right] \right\}
\slashed{\vx_{41}'}\mathcal{P}_{-} \label{gravc1}
\end{eqnarray}
The second contribution reads
\begin{eqnarray}
I^{(2)}_{graviton}=\frac{1}{18}\frac{C_{4}^2}{|\vx_{21}|^6|\vx_{31}|^8|\vx_{41}|^6}&&\int
[dw]w_{0}(t+t^2+t^3)\left\{\mathcal{P}_{+}K_{4}(w,\vx_{21}')\slashed{\vx_{21}'}U(w-\vx_{21}')\right.
\nonumber \\
 &\times& \left. \Gamma^{\mu}\overleftrightarrow{D}_{\mu}
U(w-\vx_{41}')\slashed{\vx_{41}'}K_{4}(w,\vx_{41}')\mathcal{P}_{-}\right\}
\end{eqnarray}
which can be simplified using
\begin{equation}
\Gamma^{\mu}D_{\mu}(K_{4}(w,\vx')U(w-\vx'))=-\frac{3}{2}K_{4}(w,\vx')U(w-\vx')
\end{equation}
resulting in
\begin{equation}
\label{gravc2}
I^{(2)}_{graviton}=\frac{1}{6}\frac{C_{4}^4}{|\vx_{21}|^6|\vx_{31}|^8|\vx_{41}|^6}\slashed{\vx_{21}'}\int
[dw](t+t^2+t^3)\tilde{K}_{4}(w,\vx_{21}')\tilde{K}_{4}(w,\vx_{41}')\slashed{\vx_{24}'}\slashed{\vx_{41}'}\mathcal{P}_{-}
\end{equation}
By shifting  $\vec{w}$ by $\vx_{31}'$, all integrals involved in
(\ref{gravc1}) and (\ref{gravc2}) can be expressed in terms of
$W$-functions \cite{D'Hoker:1999pj}, which are defined as
\begin{equation}
{W_{k}}^{\Delta'}(a,b)\equiv \int [dw]
\frac{w^{2\Delta'+2a+2k}_{0}}{w^{2k}}\frac{1}{(w-x)^{2\Delta'}}\frac{1}{(w-y)^{2\Delta'+2b}}
\label{Wfuncdef}
\end{equation}
and in this case, $x\equiv\vx_{23}'$ and $y=\vx_{43}'$. These are
essentially four-point contact diagrams in the inverted frame. In
order to express the amplitude by means of $W$-functions, notice that
one can replace
$(\slashed{\vec{w}}-\slashed{\vx'})K_{\Delta}(w,\vx')$ by
derivatives on $\vx'$, using the relation
\begin{equation}
\frac{(\slashed{\vec{w}}-\slashed{\vx'})}{w_{0}}\tilde{K}_{\Delta+1}(w,\vx')=\frac{1}{2\Delta}\tilde{K}_{\Delta}(w,\vx')
\label{derivslash}
\end{equation}
and then use the identity
\begin{eqnarray}
\p_{x_{i}}{W_{k}}^{\Delta'}(a,b)&=&-2x_{i}\frac{k\Delta'}{(k+\Delta'+a-2)}{W_{k+1}}^{\Delta'+1}(a-1,b-1)
\nonumber \\
&-&2(x-y)_{i}\frac{\Delta'(\Delta'+b)}{(k+\Delta'+a-2)}{W_{k}}^{\Delta'+1}(a,b)
\label{derivsWfunc}
\end{eqnarray}
There are essentially four types of $W$-functions entering the graviton exchange.
These are $W^{4}_{k}(0,0)$, $W^{4}_{k}(1,0)$, $W_{k}^{3}(2,2)$
and $\tilde{W}_{k}^{3}(2,2)$, with the tilde indicating $x \leftrightarrow y$.
Performing the derivatives and inverting the coordinates, so that
\begin{eqnarray}
\slashed{\vx'}_{21}\slashed{\vx'}_{43}\slashed{\vx'}_{41}&=&
\frac{\slashed{\vx}_{24}}{|\vx_{21}|^2|\vx_{41}|^2}+
\frac{\slashed{\vx}_{23}\slashed{\vx}_{31}\slashed{\vx}_{14}}{|\vx_{21}|^2|\vx_{31}|^2|\vx_{41}|^2} \nonumber \\
\slashed{\vx'}_{21}\slashed{\vx'}_{24}\slashed{\vx'}_{41}&=&-\frac{\slashed{\vx}_{24}}{|\vx_{21}|^2|\vx_{41}|^2} \qquad
\slashed{\vx'}_{21}\slashed{\vx'}_{23}\slashed{\vx'}_{41}=
\frac{\slashed{\vx}_{23}\slashed{\vx}_{31}\slashed{\vx}_{14}}{|\vx_{21}|^2|\vx_{31}|^2|\vx_{41}|^2}
\end{eqnarray}
and writing $\slashed{\tilde{x}}\equiv
x_{i}\gamma^{i}\mathcal{P}_{+}=x_{i}(\bar{\sigma}^{i})^{\dot{\alpha}\alpha}$
one can express the amplitude as
\begin{eqnarray}
I_{graviton}&=&-\frac{1}{6}\frac{C_{4}^4}{|\vx_{21}|^8|\vx_{31}|^8|\vx_{41}|^8}
\sum_{k=1}^{3}\left\{ \slashed{\vx}_{24} \left[ 2W_{k}^{4}(0,0)
+\frac{64}{(k+3)}W_{k}^{5}(1,0)\right. \right.\nonumber \\
&-&\left.\frac{40}{(k+3)}(\tilde{W}^{4}_{k}(2,2)+W_{k}^{4}(2,2))
+\frac{8k}{(k+3)}(W_{k+1}^{4}(1,1)-\tilde{W}_{k+1}^{4}(1,1))\right] \nonumber \\
&+&\left.\frac{\slashed{\vx}_{23}\slashed{\tilde{x}}_{31}\slashed{\vx}_{14}}{|\vx_{31}|^2}\left[\frac{16k}{(k+3)}(W_{k+1}^{4}(1,1)
-\tilde{W}_{k+1}^{4}(1,1))
\right]\right\}
\end{eqnarray}
It is possible to translate between the $W$-functions and the more
familiar $D$-functions,
$D_{\Delta_{1}\Delta_{2}\Delta_{3}\Delta_{4}}$
\cite{D'Hoker:1999pj}, which are identified with quartic scalar
interactions, and are related to the $W$-functions by inversion of
coordinates. The explicit relation in this case, is given by
\begin{equation}
{W_{k}}^{\Delta'}(a,b)=|\vx_{21}|^{2\Delta'}|\vx_{31}|^{2k}|\vx_{41}|^{2(\Delta'+b)}D_{2a-b+k,\Delta',k,\Delta'+b}
\end{equation}
One can further, rewrite these in terms of conformal ratios, $u$ and
$v$, by introducing $\bar{D}$-functions \cite{Arutyunov:2002fh},
whose properties are listed in appendix A. Using these, the
expression for the graviton exchange contribution to the
supergravity amplitude is given by
\begin{eqnarray}
I_{graviton}&=&-\frac{\pi^2}{6}
\frac{C_{4}^4}{|\vx_{13}|^8|\vx_{24}|^8} \left\{ \left[
\frac{1}{18}\bar{D}_{1414}+\frac{1}{6}\bar{D}_{2424}+\frac{1}{6}\bar{D}_{3434}+\frac{1}{12}\bar{D}_{3425}
-\frac{1}{12}\bar{D}_{4253}+\frac{1}{9}\bar{D}_{4435} \right.\right. \nonumber
\\
&-&\left.\left.\frac{1}{9}\bar{D}_{4354}+\frac{5}{72}\bar{D}_{5445}
-\frac{5}{72}\bar{D}_{4455}-\frac{1}{12}\bar{D}_{3416}-\frac{1}{12}\bar{D}_{4163}-\frac{1}{9}\bar{D}_{4426}
-\frac{1}{9}\bar{D}_{4264}
\right. \right.\nonumber \\
&-&\left.\left.\frac{5}{72}\bar{D}_{5436}-\frac{5}{72}\bar{D}_{4365}+\frac{1}{6}\bar{D}_{2426}
+\frac{2}{9}\bar{D}_{5254}+\frac{5}{36}\bar{D}_{5355}
\right]\slashed{\vx_{24}}\right.
\nonumber \\
&+&\left.\left[\frac{1}{6}\bar{D}_{3425}
-\frac{1}{6}\bar{D}_{4253}+\frac{2}{9}\bar{D}_{4435}-\frac{2}{9}\bar{D}_{4354}+\frac{5}{36}\bar{D}_{5445}
+\frac{5}{36}\bar{D}_{4455}\right]\frac{\slashed{\vx_{23}}\slashed{\tilde{x}_{31}}\slashed{\vx_{14}}}{|\vx_{13}|^2}
 \right\} \nonumber \\
\label{Gravitondirect}
\end{eqnarray}
which can be further simplified using the identities
(\ref{nontrivialid1},\ref{nontrivialid2}) (some of the manipulations are included in
appendix B).
\begin{eqnarray}
I_{graviton}&=&\frac{\pi^2}{6}
\frac{C_{4}^4}{|\vx_{13}|^8|\vx_{24}|^8} \left\{ \left[
(u+v)\left(\frac{1}{6}\bar{D}_{2525}+\frac{2}{9}\bar{D}_{3535}+\frac{5}{36}\bar{D}_{4545}\right)\right.\right. \nonumber \\
&-&\frac{1}{18}\bar{D}_{1414}-\frac{2}{9}\bar{D}_{2424}-\frac{2}{9}\bar{D}_{3434}+\frac{11}{36}\bar{D}_{4444}
-\frac{5}{72}\bar{D}_{5454}
\nonumber \\
&-&\left.\frac{1}{24}(\bar{D}_{3526}-\bar{D}_{2536})-\frac{5}{72}(\bar{D}_{4536}-\bar{D}_{3546})\right]
\slashed{\vx_{24}} \nonumber \\
&+&\left.\left[\frac{1}{12}(\bar{D}_{2536}-\bar{D}_{3526})+\frac{5}{36}(\bar{D}_{3546}-\bar{D}_{4536})\right]
\frac{\slashed{\vx_{23}}\slashed{\tilde{x}_{31}}\slashed{\vx_{14}}}{|\vx_{13}|^2}
 \right\} \nonumber \\
\label{amp1}
\end{eqnarray}
This last expression explicitly shows that the amplitude
is symmetric under exchange of $\vx_{1}$ and $\vx_{3}$, as
one should expect since
\begin{equation}
\frac{\slashed{\vx}_{23}\slashed{\tilde{x}}_{31}\slashed{\vx}_{14}}{|\vx_{13}|^2}=
-\frac{\slashed{\vx}_{21}\slashed{\tilde{x}}_{13}\slashed{\vx}_{34}}{|\vx_{13}|^2}-\slashed{\vx}_{24}
\end{equation}
It also transforms consistently under inversion,
and respects the structure of the spinor indices, given that

\subsection{Gravitino Exchange}
We will now turn to the calculation of the gravitino exchange
contribution, which is the novel part of the calculation. We
will first generalize the procedure for computing
$z$-integrals, and then we will express the result in a way that is
consistent with superconformal symmetry. We will suppress spinor
indices throughout this section, for simplicity.

Compactification of type IIB supergravity raises an interaction term between the
massless gravitino and a current of the form
\begin{equation}
\label{supcur}
\mathcal{J}^{i \mu} = \frac{1}{\tau_2} (\partial_\nu
\tau) \Gamma^\nu \Gamma^\mu \Lambda^{i *}
\end{equation}
which satisfies a conservation equation of the form \cite{Nolland:2000fx,PhysRevLett.38.1433}
\begin{equation}
\label{conserveeqn}
\Big( D_\mu - \frac{\Gamma_\mu}{2} \Big)
\mathcal{J}^{i \mu} = 0,
\end{equation}
with $D_\mu$ containing the spin connection and the Levi-Civita
connection.

One can proceed as before, by writing the solution to the equation of motion as
$\psi_{\mu}=\psi_{\mu}^{0}+\delta \psi_{\mu}$. The perturbation
can then be expressed as
\begin{equation}
\delta \psi_{\mu}(w)=\int [dz] \Theta_{\mu\nu'}(w,z)\mathcal{J}^{\nu'}(z)
\end{equation}
where $\Theta_{\mu\nu'}(w,z)$ is the bulk-to-boundary propagator for
the massless gravitino in $AdS$. Now we evaluate the action
on-shell, to obtain the integral describing the gravitino exchange.
Here one needs to be careful in taking into account the factors of
$1/2\sqrt{2}$ that precedes the current, that determine the overall
normalization. We then arrive at the expression
\begin{eqnarray}
I_{gravitino}=\frac{1}{4}\int [dw] \bar{\mathcal{J}}^{\mu}(w,\vec{x}_{2},\vec{x}_{3})
\psi_{\mu}(w,\vec{x}_{1},\vec{x}_{4}) \nonumber \\
\psi_{\mu}(w,\vec{x}_{1},\vec{x}_{4})=\int [dz] \Theta_{\mu
\nu'}(w,z)\mathcal{J}^{\nu'}(z,\vec{x}_{1},\vec{x}_{4})
\label{gravitinoex}
\end{eqnarray}
In principle, one could try to evaluate this expression using the
explicit form for the $\Theta_{\mu\nu'}(w,z)$, the bulk--to--bulk
propagator for the massless gravitino \cite{Basu:2006ti}. However,
it is simpler to generalize the method used for the graviton, to
evaluate the $z$-integral. Using translation and conformal
inversion, one can show that $\psi_{\mu}$ can be reexpressed as
\begin{equation}
\psi_{\mu}(w,\vec{x}_{1},\vec{x}_{4})=-\frac{\slashed{w}}{|w|}\frac{J_{\lambda\mu}(w)}{w^2}I_{\lambda}(w'-\vec{x}_{41}')
\frac{\slashed{\vec{x}'}_{41}}{|\vec{x}_{41}|^{6}}
\label{resgrav}
\end{equation}
where $J_{\mu\nu}(w)$ is the inversion jacobian defined before and
\begin{equation}
I_{\lambda}(z)=\int [dz]
\Theta_{\lambda\lambda'}(w,z)\Gamma^{\kappa'}\Gamma^{\lambda'}\p_{\kappa'}(z_{0}^4)
\frac{\slashed{z}}{\sqrt{z_{0}}}\left(\frac{z_{0}^4}{z^8}\right)\mathcal{P}_{+}
\end{equation}
Once again, we must write down an ansatz for $I_{\lambda}(w)$. Scale
symmetry, $d$-dimensional Poincair\'{e} symmetry and gauge
invariance \footnote{The most general ansatz one can make contains
six terms, namely, two more terms of the form
$\frac{1}{\sqrt{z_{0}}}(w_{\mu}f_{5}(t)+w_{\mu}\frac{(w.\Gamma)}{w^2}f_{6}(t))$.
However, these can be rewritten as pure gauge terms, and so can be
removed.} suggests
\begin{equation}
I_{\mu}(w)=\frac{1}{\sqrt{z_{0}}}\Big(\delta_{\mu
0}f_{1}(t)+\gamma_{\mu}f_{2}(t)+\frac{(w.\Gamma)}{w^2}\delta_{\mu
0}f_{3}(t)+\gamma_{\mu}\frac{(w.\Gamma)}{w^2}f_{4}(t)\Big)P_{+}
\label{ansatz}
\end{equation}
where $t=w_{0}^2/w^2$. The next step is to use the Rarita-Schwinger
Green's function equation for $\Theta_{\lambda\lambda'}(w,z)$ to
find an equation for $I_{\lambda}(w)$, namely
\begin{equation}
{W_{\mu}}^{\rho}I_{\rho}(w)=\left(\Gamma_{\mu}\Gamma^{\nu}\Gamma^{\rho}D_{\nu}-\Gamma^{\rho}D_{\nu}-\Gamma_{\mu}D^{\rho}
+\Gamma^{\nu}D_{\nu}\delta_{\mu}^{\rho}
+\frac{3}{2}\Gamma_{\mu}\Gamma^{\rho}-\frac{3}{2}\delta_{\mu}^{\rho}\right)I_{\rho}(w)=\mathcal{\tilde{J}}_{\mu}(w)
\label{greenfunc}
\end{equation}
with
\begin{eqnarray}
\mathcal{\tilde{J}}_{\mu}(w)&=&\int [dz] \delta(z-w)g_{\mu\rho'}\Gamma^{\nu}\Gamma^{\rho'}\p_{\nu}(z_{0}^{4})
\frac{\slashed{z}}{\sqrt{z_{0}}}
\left(\frac{z_{0}}{z^2}\right)^{4}\mathcal{P}_{+} \nonumber \\
&=&\frac{4t^{3}}{\sqrt{w_{0}}}\left\{2\delta_{\mu 0}\frac{(w\cdot \Gamma)}{w^2}
-2\gamma_{\mu}t+\gamma_{\mu}\frac{(w\cdot \Gamma)}{w^2}\right\}
\mathcal{P}_{+}
\end{eqnarray}
Needless to say that the application of the wave operator is quite tedious.
We simply give the results of these calculations. The left hand
side of (\ref{greenfunc}) reads
\begin{eqnarray}
{W_{\mu}}^{\rho}\left[ \frac{\delta_{\rho 0}}{\sqrt{w_{0}}}f_{1}(t) \right]\mathcal{P}_{+}
&=&\frac{1}{\sqrt{w_{0}}}\left[ 2tf_{1}'(t)\left(w_{0}\frac{w_{\mu}}{w^2}-\delta_{\mu 0}\frac{(w\cdot \Gamma)}{w^2}
-\gamma_{\mu}\frac{(w\cdot \Gamma)}{w^2}\right)\right.
\nonumber \\
&+&\left.(2t^{2}f_{1}'(t)+3f_{1}(t))\gamma_{\mu}-3f_{1}(t)\delta_{\mu 0}\right]\mathcal{P}_{+}
\nonumber \\
{W_{\mu}}^{\rho}\left[ \frac{\gamma_{\rho}}{\sqrt{w_{0}}}f_{2}(t) \right]\mathcal{P}_{+}
&=&\frac{1}{\sqrt{w_{0}}}\left[ (6tf_{2}'(t)+3f_{2}(t))\left(\gamma_{\mu}-\delta_{\mu 0}\right) \right.
\nonumber \\
&+&\left. 6tf_{2}'(t)\left(w_{0}\frac{w_{\mu}}{w^{2}}-\gamma_{\mu}\right)
\frac{(w\cdot \Gamma)}{w^{2}}\right]\mathcal{P}_{+}
\nonumber \\
{W_{\mu}}^{\rho}\left[ \frac{\delta_{\rho 0}}{\sqrt{w_{0}}}\frac{(w\cdot \Gamma)}{w^{2}}f_{3}(t) \right]\mathcal{P}_{+}
&=&\frac{1}{\sqrt{w_{0}}}\left[(2t^{2}f_{3}'(t)+2tf_{3}(t))\left(\gamma_{\mu}
-\delta_{\mu 0}-\gamma_{\mu}\frac{(w\cdot \Gamma)}{w^{2}}
+2w_{0}\frac{w_{\mu}}{w^{2}}\right) \right.
\nonumber \\
&-&\left.(2tf_{3}'(t)+2f_{3}(t))w_{0}w_{\mu}\frac{(w\cdot \Gamma)}{w^{4}}\right]\mathcal{P}_{+}
\nonumber \\
{W_{\mu}}^{\rho}\left[ \frac{\gamma_{\rho}}{\sqrt{w_{0}}}\frac{(w\cdot \Gamma)}{w^{2}}f_{4}(t) \right]\mathcal{P}_{+}
&=&\frac{1}{\sqrt{w_{0}}}\left[(6tf_{4}'(t)+6f_{4}(t))\left(t\gamma_{\mu}-\delta_{\mu 0}\frac{(w\cdot \Gamma)}{w^{2}}
+w_{0}w_{\mu}\frac{(w\cdot\Gamma)}{w^{4}}\right)\right.
\nonumber \\
&-&\left.(6tf_{4}'(t)-6f_{4}(t))\gamma_{\mu}\frac{(w\cdot\Gamma)}{w^{2}}\right]\mathcal{P}_{+}
\end{eqnarray}
So substitution of the ansatz
(\ref{ansatz}) gives a set of equations for the undetermined
coefficients $f_{i}(t)$. The system, however, is overdetermined, as
it has 6 equations and 4 unknowns:
\begin{eqnarray}
-3f_{1}(t)-6tf_{2}'(t)-3f_{2}(t)-2t^2f_{3}'(t)-2tf_{3}(t)&=&0
\nonumber \\
2t^2f_{1}'(t)+3f_{1}(t)+6tf_{2}'(t)+3f_{2}(t)+2t^2f_{3}'(t)+2tf_{3}(t)+6t^2f_{4}'(t)+6tf_{4}(t)&=&-8t^4
\nonumber \\
-2f_{1}'(t)-6tf_{2}'(t)-2t^2f_{3}'(t)-2tf_{3}(t)-6tf_{4}'(t)+6f_{4}(t)&=&4t^3
\nonumber \\
2tf_{1}'(t)+6tf_{2}'(t)+4t^2f_{3}'(t)+4tf_{3}(t)&=&0
\nonumber \\
-2tf_{1}'(t)-6tf_{4}'(t)-6f_{4}(t)&=&8t^3
\nonumber \\
-2tf_{3}'(t)-2f_{3}(t)+6tf_{4}'(t)+6f_{4}(t)&=&0
\end{eqnarray}
One must then look for a consistent solution. In this case it is
given by
\begin{eqnarray}
f_{1}(t)&=&-\frac{1}{3}(2t+3t^2+4t^3) \qquad
f_{2}(t)=\frac{1}{9}(2t+t^2) \nonumber \\
f_{3}(t)&=&\frac{1}{3}(t+2t^2) \hspace{23.mm}
f_{4}(t)=\frac{1}{9}(t+2t^2) \label{consistentsol}
\end{eqnarray}
The functions above are regular on $t=1$, as expected, and vanish
when $t$ approaches zero. Using these results on (\ref{resgrav}),
one may compute $I_{gravitino}$. Upon inversion of the current
$\bar{\mathcal{J}}^{\mu}$, the amplitude takes the form
\begin{equation}
I_{gravitino}=\frac{1}{4}\frac{C_{4}^4\slashed{\vx'}_{21}}{|\vx_{21}|^6|\vx_{31}|^8|\vx_{41}|^6}\int
[dw]
\mathcal{P}_{-}\tilde{K}_{7/2}(w,\vx_{21}')\Gamma^{\lambda}\Gamma^{\nu}\p_{\nu}(\tilde{K}_{4}(w,\vx_{31}'))
I_{\lambda}(w-\vx_{41}')\slashed{\vx'}_{41}
\label{gravitinoex2}
\end{equation}
Substituting (\ref{ansatz}) with (\ref{consistentsol}), the
resulting expression can be split into two contributions. To see
this, we can rewrite the quantity between the chiral projectors, as
\begin{eqnarray}
&&\mathcal{P}_{-}\tilde{K}_{7/2}(w,\vx_{21}')\Gamma^{\mu}\Gamma^{\nu}\p_{\nu}\tilde{K}_{4}(w,\vx_{31}')I_{\mu}(w-\vx_{41}')
=\mathcal{P}_{-}\tilde{K}_{4}(w,\vx_{21}')(\slashed{w}-\slashed{\vx}_{21}')\Big[ \left(f_{1}(t)\gamma_{0}\right.
\nonumber \\
&&\left.\left.-3f_{2}(t)\right)\Gamma^{\nu}\p_{\nu}\tilde{K}_{4}(w,\vx_{31}')
-6tf_{4}(t)\mathcal{P}_{-}\Gamma^{\nu}\p_{\nu}\tilde{K}_{4}(w,\vx_{31}')(\slashed{w}-\slashed{\vx}_{41}') \right]
\end{eqnarray}
The first contribution involves $f_{1}(t)$ and $f_{2}(t)$, whereas the second contribution is proportional to $f_{4}(t)$.
We can simplify further by using the identity
\begin{equation}
\Gamma^{\mu}D_{\mu}\tilde{K}_{4}(w,\vx')=4\gamma_{0}\tilde{K}_{4}(w,\vx')-8(\slashed{w}-\slashed{\vx'})\tilde{K}_{5}(w,\vx')
\end{equation}
so doing some algebra and simplifying, one gets
\begin{equation}
I_{gravitino}^{(1)}=-\frac{1}{3}\frac{C_{4}^4\slashed{\vx_{21}'}}{|\vx_{21}|^{6}|\vx_{31}|^{8}|\vx_{41}|^{6}}
\int [dw]
\tilde{K}_{4}(w,\vx_{21}')w_{0}\p_{0}\tilde{K}_{4}(w,\vx_{31}')
(t+t^2+t^3)(\slashed{\vec{w}}-\slashed{\vx_{21}'})\slashed{\vx_{41}'}\mathcal{P}_{-}
\end{equation}
\begin{equation}
I_{gravitino}^{(2)}=-\frac{1}{6}\frac{C_{4}^4\slashed{\vx_{21}'}}{|\vx_{21}|^{6}|\vx_{31}|^{8}|\vx_{41}|^{6}}
\int [dw]
\tilde{K}_{4}(w,\vx_{21}')w_{0}\p_{0}\tilde{K}_{4}(w,\vx_{31}')
(t^2+2t^3)(\slashed{\vec{w}}-\slashed{\vx_{41}'})\slashed{\vx_{41}'}\mathcal{P}_{-}
\end{equation}
The resulting integrals have the same form as those that occurred in
the graviton case. Hence, we can make use of the same tricks. First
we turn all terms of the form $(\slashed{\vec{w}}-\slashed{\vx'})$
into derivatives using (\ref{derivslash}), and then group in terms
of $W$-functions, which were defined in (\ref{Wfuncdef}), but taking
$x\equiv \vx_{24}'$ and $y\equiv \vx_{34}'$. Then
$I_{gravitino}^{(1)}$ can be rewritten in terms of $W_{k}^{3}(1,1)$
and $W_{k}^{3}(2,2)$, for $k=1,2,3$, and $I_{gravitino}^{(2)}$ in
terms of  $W_{k}^{4}(1,0)$ and $W_{k}^{4}(2,1)$, for $k=1,2$.
Finally, one simplifies by doing the derivatives using
(\ref{derivsWfunc}) and inverting the coordinates as before, so
$I_{gravitino}$ is given by
\begin{eqnarray}
I_{gravitino}&=&-\frac{1}{3}\frac{C_{4}^4}{|\vx_{21}|^{8}|\vx_{31}|^{8}|\vx_{41}|^{8}} \left[
\sum_{k=1}^{3}\left\{ \slashed{\vx}_{24}\left[ \frac{4k}{(k+2)}W_{k+1}^{4}(0,0)
-\frac{8k}{(k+3)}W_{k+1}^{4}(1,1)\right]\right.\right.
\nonumber \\
&+&\left.\frac{\slashed{\vx}_{23}\slashed{\tilde{x}}_{31}\slashed{\vx}_{14}}{|\vx_{31}|^{2}}\left[-\frac{16}{(k+2)}W_{k}^{4}(1,1)
+\frac{40}{(k+3)}W_{k}^{4}(2,2) \right] \right\}+\sum_{k=1}^{2}\left\{ \slashed{\vx}_{24}\left[
-\frac{8k}{(k+3)}\times\right.\right.
\nonumber \\
&+&\left.\left.(\tilde{W}_{k+1}^{4}(1,1)+W_{k+1}^{4}(1,1))+\frac{2k}{(k+3)}(8W_{k+1}^{5}(1,0)+10W_{k+1}^{4}(2,2))\right] \right.
\nonumber \\
&+&\left.\left.\frac{\slashed{\vx}_{23}\slashed{\tilde{x}}_{31}\slashed{\vx}_{14}}{|\vx_{31}|^{2}}\left[
-\frac{8k}{(k+3)}W_{k+1}^{4}(1,1)
+\frac{20k}{(k+4)}W_{k+1}^{4}(2,2)\right]\right\}\right]
\end{eqnarray}
Again, it is possible to translate between $W$-functions
and $D$-functions. In this case, the relation is
\begin{equation}
W_{k}^{\Delta'}(a,b)=|\vx_{21}|^{2\Delta'}|\vx_{31}|^{2(\Delta'+b)}|\vx_{41}|^{2k}D_{2a-b+k,\Delta',\Delta'+b,k}
\end{equation}
From here it is straightforward to rewrite the gravitino exchange diagram in terms of conformal invariant rations, by
introducing the $\bar{D}$-functions. The amplitude then reads
\begin{eqnarray}
I_{gravitino}&=&-\frac{\pi^2}{3}\frac{C^{4}_{4}}{|\vx_{13}|^8|\vx_{24}|^8}\left\{\left[\frac{1}{9}\bar{D}_{2442}
+\frac{1}{6}\bar{D}_{3443}
+\frac{1}{9}\bar{D}_{4444}-\frac{1}{6}\bar{D}_{3452}-\frac{2}{9}\bar{D}_{4453}\right.\right.
\nonumber \\
&-&\frac{5}{72}\bar{D}_{5454}+\frac{1}{18}\bar{D}_{4462}
+\frac{5}{72}\bar{D}_{5463}-\frac{1}{12}\bar{D}_{2453}-\frac{1}{9}\bar{D}_{3452}+\frac{1}{18}\bar{D}_{3463}
\nonumber \\
&+&\left.\frac{5}{72}\bar{D}_{4464}\right]\slashed{\vx_{24}}+\left[-\frac{1}{9}\bar{D}_{2451}
-\frac{1}{4}\bar{D}_{3452}-\frac{2}{9}\bar{D}_{4453}
+\frac{1}{12}\bar{D}_{3461}\right.
\nonumber \\
&+&\left.\left.\frac{1}{6}\bar{D}_{4462}+\frac{5}{36}\bar{D}_{5463}\right]
\frac{\slashed{\vx_{23}}\slashed{\tilde{x}_{31}}\slashed{\vx_{14}}}{|\vx_{13}|^2}
\right\}
\label{Gravitinodirect}
\end{eqnarray}
This expression can be shortened by using the identities of appendix
A, as described in appendix B. The final expression one is left with
is then
\begin{eqnarray}
I_{gravitino}&=&\frac{\pi^2}{3}\frac{C^{4}_{4}}{|\vx_{13}|^8|\vx_{24}|^8}\left\{\left[\frac{1}{18}\bar{D}_{2442}
+\frac{1}{12}\bar{D}_{3443}
-\frac{2}{9}\bar{D}_{4444}-\frac{1}{12}\bar{D}_{3452}-\frac{1}{9}\bar{D}_{4453}\right.\right.
\nonumber \\
&+&\left.\frac{5}{36}\bar{D}_{5454}\right]\slashed{\vx}_{24}+\left[\frac{1}{36}\bar{D}_{2451}
+\frac{1}{12}\bar{D}_{3452}+\frac{1}{12}\bar{D}_{4453}-\frac{1}{12}\bar{D}_{2552}-\frac{1}{6}\bar{D}_{3553}\right.
\nonumber \\
&-&\left.\left.\frac{5}{36}\bar{D}_{4554}\right]
\frac{\slashed{\vx_{23}}\slashed{\tilde{x}_{31}}\slashed{\vx_{14}}}{|\vx_{13}|^2}
\right\}
\label{amp2}
\end{eqnarray}

\subsection{Quartic Diagram}
The last diagram that one needs to compute is the quartic
interaction. The necessary couplings are obtained from the cubic
vertices involving two dilatinos and $P_{\mu}$, by considering
variations of the dilaton factor $\tau_{2}^{-1}$. The relevant terms
in the action become
\begin{equation}
I_{quartic}=\int [dz]
\left[\frac{3}{8}\bar{\hat{\Lambda}}\Gamma^{\mu}\p_{\mu}\left(\frac{i\delta
\tau}{\sqrt{2}\tau_{2}}\right)\left(-\frac{i\delta
\bar{\tau}}{\sqrt{2}\tau_{2}}\right)\hat{\Lambda}-\frac{3}{8}\bar{\hat{\Lambda}}\Gamma^{\mu}\p_{\mu}\left(-\frac{i\delta
\bar{\tau}}{\sqrt{2}\tau_{2}}\right)\left(\frac{i\delta
\tau}{\sqrt{2}\tau_{2}}\right)\hat{\Lambda}\right]
\end{equation}
We now proceed as before. One replaces the bulk fields in terms of
their boundary values. There will be two contributions to the
diagram, coming from interchange of the fields in $x_{1}$ and
$x_{3}$. We compute explicitly one of the contributions, and obtain
the second by interchange of $\vx_{1}$ and $\vx_{3}$, and some
simple manipulations involving the $\bar{D}$-functions and
$\Gamma$-matrices.

The integral arising from the action has the structure
\begin{equation}
\int [dw]
\mathcal{P}_{+}U(w-\vx_{2})K_{4}(w,\vx_{2})\Gamma^{\mu}\p_{\mu}K_{4}(w,\vx_{3})
K_{4}(w,\vx_{1})K_{4}(w,\vx_{4})U(w-\vx_{4})\mathcal{P}_{-}
\label{quarticint}
\end{equation}
Using translation invariance we can set $\vx_{4}\to 0$. Furthermore,
one can invert the expression to make the integrand simpler
\begin{equation}
\frac{-\slashed{\vx_{24}'}C_{4}^{4}}{|\vx_{24}'|^{-6}|\vx_{34}'|^{-8}|\vx_{14}'|^{-8}}
\int [dw]
\mathcal{P}_{-}\tilde{K}_{4}(w,\vx_{24}')\tilde{K}_{4}(w,\vx_{14}')U(w-\vx_{24}')
\Gamma^{\mu}\p_{\mu}\tilde{K}_{4}(w,\vx_{34}')
\frac{w_{0}^{4}}{\sqrt{w_{0}}}\mathcal{P}_{-}
\end{equation}
The integrand can be simplified by working out the gamma-matrix
algebra, by using similar manipulations as the ones performed in
previous sections. It is then simple to rewrite this integral as a
sum of a single gamma-matrix term, and a triple-gamma matrix term
\begin{equation}
\frac{
\slashed{\vx_{24}'}C_{4}^{4}}{|\vx_{24}'|^{-6}|\vx_{34}'|^{-8}|\vx_{14}'|^{-8}}
\int [dw]
\tilde{K}_{4}(w,\vx_{24}')\tilde{K}_{4}(w,\vx_{14}')w_{0}^{4}\left\{4
+\slashed{\vx_{32}'}\slashed{\p}_{34'}
\right\}\tilde{K}_{4}(w,\vx_{34}')\mathcal{P}_{-}
\end{equation}
The integrals here can be again turn into $W$-functions by
translating $\vec{w}$ by $\vx_{14}'$, and using $x=\vx_{21}'$,
$y=\vx_{31}'$, in the definition. Both terms are given by
$W_{4}^{4}(0,0)$, but the action of the derivative in the second
term, yields additional terms of the form
\begin{equation}
\slashed{\p}_{34'}W_{4}^{4}(0,0)=-\frac{16}{3}\left(W_{4}^{5}(0,0)\slashed{\vx'}_{32}+W_{5}^{4}(0,1)\slashed{\vx'}_{31}\right)
\end{equation}
The next step is to invert back to the original set of coordinates.
The $W$-integrals in this case are given in terms of $D$-integrals with $x$-dependence of the form
\begin{equation}
W_{k}^{\Delta'}(a,b)=|\vx_{14}|^{2k}|\vx_{24}|^{2\Delta'}|\vx_{34}|^{2(\Delta'+b)}D_{k,\Delta',\Delta'+b,2a-b+k}
\end{equation}
so (\ref{quarticint}) is then given by
\begin{eqnarray}
&&4C_{4}^{4}\left[\slashed{\vx'}_{24}\left(D_{4444}-\frac{4}{3}\right)|\vx_{23}|^{2}D_{4554}
+\slashed{\vx}_{23}\slashed{\vx}_{31}\slashed{\vx}_{14}D_{5454}\right]
\nonumber \\
&=&\frac{5}{54}\frac{\pi^{2}C_{4}^{4}}{|\vx_{13}|^{8}|\vx_{24}|^{8}}\left[\slashed{\vx'}_{24}(2\bar{D}_{4444}-\bar{D}_{5445})+
\frac{\slashed{\vx}_{23}\slashed{\vx}_{31}\slashed{\vx}_{14}}{|\vx_{13}|^{2}}\bar{D}_{5454}\right]
\end{eqnarray}
where in the last line we introduced the previously defined
$\bar{D}$-functions.

The second contribution to the diagram is given by interchange of
$\vx_{1}$ and $\vx_{3}$. In terms of $\bar{D}$-functions, one has
\begin{equation}
\frac{5}{54}\frac{\pi^{2}C_{4}^{4}}{|\vx_{13}|^{8}|\vx_{24}|^{8}}\left[\slashed{\vx'}_{24}(2\bar{D}_{4444}
-\bar{D}_{4455}-\bar{D}_{5454})-
\frac{\slashed{\vx}_{23}\slashed{\vx}_{31}\slashed{\vx}_{14}}{|\vx_{13}|^{2}}\bar{D}_{5454}\right]
\end{equation}
Finally, the total contribution to the contact diagram is then given
by
\begin{equation}
I_{quartic}=\frac{5}{72}\frac{\pi^{2}C_{4}^{4}}{|\vx_{13}|^{8}|\vx_{24}|^{8}}\left[\slashed{\vx'}_{24}(\bar{D}_{5445}
-2\bar{D}_{4444})-
\frac{\slashed{\vx}_{23}\slashed{\tilde{x}}_{31}\slashed{\vx}_{14}}{|\vx_{13}|^{2}}\bar{D}_{5454}\right]
\label{amp3}
\end{equation}
where the identity
$4\bar{D}_{4444}=\bar{D}_{5445}+\bar{D}_{4455}+\bar{D}_{5454}$ has
been used.
\subsection{The sum of the three contributions}
The final result is then given by adding the contributions from
(\ref{amp1}), (\ref{amp2}) and (\ref{amp3}). However, we still need
to take into account the normalization of the quadratic action. Two--point
functions are given by \cite{Freedman:1998tz,Muck:1998iz}
\begin{eqnarray}
\langle
\Ot(\vx_{1})\Obt(\vx_{2})\rangle&=&\frac{4C_{4}}{2\kappa_{5}^{2}}\frac{1}{|\vx_{12}|^{8}}=
\frac{3N^{2}}{\pi^{4}}\frac{1}{|\vx_{12}|^{8}}
\nonumber \\
\langle
\Lambda(\vx_{1})\bar{\Lambda}(\vx_{2})\rangle&=&\frac{C_{4}}{2\kappa_{5}^{2}}\frac{\slashed{\vx}_{12}}{|\vx_{12}|^{8}}
=\frac{3N^{2}}{4\pi^{4}}\frac{\slashed{x}_{12}}{|\vx_{12}|^{8}}
\end{eqnarray}
having used $2\kappa_{5}^{2}=\frac{8\pi^2}{N^2}$. We define normalized
operators as $\mathcal{\tilde{O}}_{\tau}=\xi_{1}\Ot$ and
$\tilde{\Lambda}=\xi_{2}\Lambda$, such that the two--point functions give
\begin{equation}
\langle \mathcal{\tilde{O}}_{\tau}(\vx_{1})\tilde{\bar{\mathcal{O}}}_{\tau}(\vx_{2})\rangle=\frac{1}{|\vx_{12}|^{8}} \qquad
\langle \tilde{\Lambda}(\vx_{1})\bar{\tilde{\Lambda}}(\vx_{2})\rangle=\frac{\slashed{\vx}_{12}}{|\vx_{12}|^{8}}
\end{equation}
so the normalization constants are then
\begin{equation}
\xi_{1}= \frac{\pi^2}{\sqrt{3}N} \qquad
\xi_{2}=\frac{2}{\sqrt{3}}\frac{\pi^2}{N}
\end{equation}
Finally, the overall normalization constant for the connected
contribution to the four-point function, is given by
$(2\kappa_{5}^{2})^{-1}\xi_{1}^{2}\xi_{2}^{2}=\frac{\pi^{6}}{18N^2}$.
We can then recast the full contribution to the connected part of
the four-point function as follows
\begin{eqnarray}
&&\langle\mathcal{\tilde{O}}_{\tau}(\vx_{1})\tilde{\Lambda}(\vx_{2})
\mathcal{\tilde{O}}_{\bar{\tau}}(\vx_{3})\bar{\tilde{\Lambda}}(\vx_{4})\rangle_{conn}=
\frac{1}{3N^2}\frac{1}{|\vx_{13}|^8|\vx_{24}|^8}\left\{ \left[
(u+v)\left(6\bar{D}_{2525}+8\bar{D}_{3535}\right.\right.\right.
\nonumber \\
&&+\left.5\bar{D}_{4545}\right)-2\bar{D}_{1414}-8\bar{D}_{2424}-8\bar{D}_{3434}-35\bar{D}_{4444}+\frac{15}{2}\bar{D}_{5454}
+4\bar{D}_{2442}+6\bar{D}_{3443}
\nonumber \\
&&+\left.15\bar{D}_{5445}-6\bar{D}_{3452}-8\bar{D}_{4453}
+\frac{3}{2}(\bar{D}_{2536}-\bar{D}_{3526})+\frac{5}{2}(\bar{D}_{3546}-\bar{D}_{4536})\right]\slashed{\vx_{24}}
\nonumber \\
&&+\left[3(\bar{D}_{2536}-\bar{D}_{3526})+5(\bar{D}_{3546}-\bar{D}_{4536})+2\bar{D}_{2451}
+6\bar{D}_{3452}+6\bar{D}_{4453}-6\bar{D}_{2552}\right.
\nonumber \\
&&\left.\left.-12\bar{D}_{3553}-10\bar{D}_{4554}-15\bar{D}_{5454}\right]
\frac{\slashed{\vx_{23}}\slashed{\tilde{x}_{31}}\slashed{\vx_{14}}}{|\vx_{13}|^2}
 \right\}
\label{finalamp}
\end{eqnarray}
\section{The OPE Interpretation}
In this section we analyze the results of the 4-point function by
using the Operator Product Expansion (OPE) interpretation. As
mentioned in the introduction, the key idea is to assume that the
amplitude can be expressed as a double OPE expansion of the form
\begin{equation}
\langle
\mathcal{O}(\vx_{1})\mathcal{O}(\vx_{2})\mathcal{O}(\vx_{3})\mathcal{O}(\vx_{4})\rangle=\sum_{m,n}
\frac{{C^{m}}_{13}(\vx_{13},\p_{1}){C^{n}}_{24}(\vx_{24},\p_{2})
\langle
\mathcal{O}_{m}(\vx_{1})\mathcal{O}_{n}(\vx_{2})\rangle}{|\vx_{13}|^{\Delta_{1}+\Delta_{3}-\Delta_{m}}|\vx_{24}|^{\Delta_{2}+\Delta_{4}-\Delta_{n}}}
\end{equation}
in the limit in which $\vx_{13}\rightarrow 0$ and
$\vx_{24}\rightarrow 0$. The operator algebra structure constants $C_{ij}^{m}(x,\p)$
can be expanded in a power series, and for primary operators
$\mathcal{O}_{n}$, these are fixed by their conformal dimensions and by
their ratios of three-point function and the two-point function
normalization constants, $C_{ijk}/C_{k}$, with $C_{ijk}=C^{k}_{ij}(0,0)$
and $C_{k}\sim\langle \mathcal{O}_{i}\mathcal{O}_{k}\rangle$.

The above OPE's receives contributions from the set of primary
operators and their conformal descendents, some of which are not
protected under quantum corrections. By the AdS/CFT correspondence,
one can identify such operators as belonging to two separate
classes: long operators, which are dual to string states, and
multi-trace operators, which can be obtained by normal ordering
products of single trace operators, and which are dual to
multi-particle states.

Examples studied in the literature have shown that in the OPE
interpretation of supergravity amplitudes, all contributions from
long operators decouple, as they acquire large anomalous dimensions
as $\lambda\rightarrow \infty$, so string states decouple
consistently \cite{D'Hoker:1999jp}. On the other hand, the
asymptotic expansion of the amplitude will contain singularities,
which exactly match the contributions of the conformal blocks to the
OPE. This is, there is a correspondence between exchange
supergravity diagrams and the contribution to the OPE from the dual
operator (short and protected) and its descendents
\cite{D'Hoker:1999jp}. We verify explicitly that the same behaviour
holds for our particular calculation.

Supergravity amplitudes also have shown to contain logarithmic terms
in their asymptotic expansions, which are to be interpreted as
renormalization effects coming from double-trace operators produced
in the OPE of two short operators, as pointed out by Witten. It is
easy to see that double-trace operators will receive anomalous
dimensions of order $O(\frac{1}{N^2})$, and careful consideration of
the terms, allows for the computation of the anomalous dimensions of
these operators \cite{Symanzik:1971vw}. We will also explicitly
compute the leading order log terms for the different channels. This
analysis is particularly clear for the $s$-channel,
$\vx_{12}\rightarrow 0$, $\vx_{34}\to 0$, given that the log terms
are precisely the leading order singularities.

To analyze the supergravity amplitude, we start by obtaining the
asymptotic expansions of the four-point function on the different
channels. The result should be then compared to the contributions
coming from the primary operators to the OPE (the conformal partial
waves). We restrict to the contributions coming from the
stress-energy tensor on the $t$-channel, and the supercurrent on the
$u$-channel. Since there are no available short-distance expansions
of conformal partial wave amplitudes of half-integer operators in
the literature, we will compute such contributions using free-field
theory. However this task can be generalized to include the
contributions from higher spin operators and descendents, and a
useful reference to start is the work on the OPE of two spin 1/2
particles by Dobrev, et. al. \cite{Dobrev:1973zg}.
\subsection{Free Field Theory OPE's}
In the free field limit, Yang-Mills theory reduces to the
abelian theory (apart from the irrelevan dependence on $N$).
In the abelian theory, the operators belonging to the current
multiplet which are dual to the 5d massless dilaton-axion and
dilatino, of Type IIB supergravity compactified in $AdS_{5}\times
S^{5}$, are given by \cite{Bergshoeff:1980is}
\begin{equation}
\mathcal{O}_{\tau}=\frac{1}{g_{YM}^{2}}(F_{\mu\nu}^{-})^2
\hspace{10.mm}
\Lambda_{\alpha}=-\frac{1}{g_{YM}^{2}}{\sigma^{\mu\nu}}_{\alpha}^{\beta}F_{\mu\nu}^{-}\lambda_{\beta}
\end{equation}
where $F^{\pm}=\frac{1}{2}(F\pm \tilde{F})$ are the (anti)self-dual
components of the field strength and
$\tilde{F}_{\mu\nu}=\frac{i}{2}\epsilon_{\mu\nu\rho\sigma}F^{\rho\sigma}$,
so $\bar{\sigma}^{\mu\nu}$ is self-dual and $\sigma^{\mu\nu}$ is
anti-self dual\footnote{Notation for Weyl spinors follows \cite{Lykken:1996}}.
One can compute the OPE's of two current multiplet
operators by using Wick's theorem and the following propagators
\begin{eqnarray}
\langle F^{+}_{\mu\nu}(\vx_{1})F^{-}_{\rho\sigma}(\vx_{2})
\rangle&=&\frac{g_{YM}^2}{8\pi^2}(\eta_{\nu\sigma}\p^{1}_{\mu}\p^{2}_{\rho}-\eta_{\nu\rho}\p^{1}_{\mu}\p^{2}_{\sigma}
-\eta_{\mu\sigma}\p^{1}_{\nu}\p^{2}_{\rho}
+\eta_{\mu\rho}\p^{1}_{\nu}\p^{2}_{\sigma})
\frac{1}{|\vx_{12}|^2} \nonumber \\
\langle
\lambda_{\alpha}(\vx_{1})\bar{\lambda}_{\dot{\alpha}}(\vx_{2})\rangle&=&
-\frac{g_{YM}^2}{4\pi^2}(\sigma^{\mu})_{\alpha\dot{\alpha}}\p_{\mu}\frac{1}{|\vx_{12}|^2}
\end{eqnarray}
The quantities $\langle F^{-}_{\mu\nu}(\vx_{1}) F^{-}_{\rho\sigma}(\vx_{2})\rangle$ and
$\langle F^{+}_{\mu\nu}(\vx_{1}) F^{+}_{\rho\sigma}(\vx_{2})\rangle$
vanish for separate points, as they only give raise to contact
terms.
We will also require the (on-shell) free-field theory expressions
for the energy-momentum tensor $T_{\mu\nu}$
\footnote{Here we have only included the contribution to $T_{\mu\nu}$
coming from the gauge field, given that naively, it is the only term that will be relevant
in the free field limit, but one should be careful with the normalization
between the different contributions from scalars, $T_{\mu\nu}^S$, fermions, $T_{\mu\nu}^{F}$
and gauge fields, $T_{\mu\nu}^{V}$ as it is discussed in \cite{Osborn:1993cr}. We will drop the
index $V$ for convenience from this point forward.},
and the supercurrent $\Sigma_{\nu\alpha}$
\begin{eqnarray}
T_{\mu\nu}^{V}&=&\frac{1}{2g_{YM}^2}[\eta_{\mu\nu}(F^{-}_{\rho\sigma})^2-4{F^{-\mu}}_{\rho}F^{-\nu\rho} + h.c.] \\
{\Sigma^{\mu}}_{\alpha}&=&-\frac{1}{g_{YM}^2}(\sigma^{\kappa\nu}
F_{\kappa\nu}^{-}\sigma^{\mu})_{\alpha\dot{\alpha}}\bar{\lambda}^{\dot{\alpha}}
\end{eqnarray}
Using these expressions we can determine the overall form of the OPE
\begin{eqnarray}
\mathcal{O}_{\tau}(\vx_{1})\Obt(\vx_{2})&\sim&
\frac{4}{\pi^2}
\frac{{x_{12}}^{\mu}{x_{12}}^{\nu}}{{|\vx_{12}|}^6}T_{\mu\nu} \\
\Lambda_{\alpha}(\vx_{1})\bar{\Lambda}_{\dot{\alpha}}(\vx_{2})&\sim&
\frac{1}{\pi^2}\frac{{x_{12}}^{\mu}(\sigma^{\nu})_{\alpha\dot{\alpha}}}{{|\vx_{12}|}^4}T_{\mu\nu}
\\
\mathcal{O}_{\tau}(\vx_{1})\bar{\Lambda}_{\dot{\alpha}}(\vx_{2})&\sim&
\frac{2}{\pi^2}\frac{{x_{12}}^{\mu}{x_{12}}^{\nu}}{{|\vx_{12}|}^6}{(\bar{\epsilon}
\bar{\sigma}_{\mu})_{\dot{\alpha}}}^{\alpha}\Sigma_{\nu \alpha}
\end{eqnarray}
Now one can use these OPEs to compute the contribution
from these operators to the four-point function. Therefore,  we will also need to
determine their two-point functions. A straightforward computation shows that
\begin{eqnarray}
\langle
T_{\mu\nu}(\vx_{1})T_{\rho\sigma}(\vx_{2})\rangle&=&\frac{4}{\pi^4}\left\{J_{\mu\rho}(\vx_{12})J_{\nu\sigma}(\vx_{12})
+J_{\mu\sigma}(\vx_{12})J_{\nu\rho}(\vx_{12})-\frac{1}{2}
\eta_{\mu\nu}\eta_{\rho\sigma}\right\}\frac{1}{{|\vx_{12}|}^8} \nonumber
\\
\langle
{\Sigma^{\mu}}_{\alpha}(\vx_{1}){\bar{\Sigma}^{\nu}}_{\dot{\alpha}}(\vx_{2})\rangle
&=&\frac{2}{\pi^4}\left\{\slashed{\vx_{12}}J^{\mu\nu}(\vx_{12})
+\frac{1}{2}\sigma^{\mu}{x_{12}}^{\nu}-\frac{1}{4}\sigma^{\mu}\bar{\sigma}^{\nu}
\slashed{\vx_{12}}\right\}_{\alpha\dot{\alpha}}\frac{1}{{|\vx_{12}|}^8}
\end{eqnarray}
where $J_{\mu\nu}$ is the conformal jacobian, which was previously
defined. Using these results, we can now compute the contributions
to the four-point function, coming from the previous conformal
blocks. For the $t$-channel, $\vx_{13}\rightarrow0$ and
$\vx_{24}\rightarrow0$, so the short--distance expansion has the form
\begin{eqnarray}
\langle
\mathcal{O}_{\tau}(\vx_{1})\Lambda_{\alpha}(\vx_{2})\Obt(\vx_{3})\bar{\Lambda}_{\dot{\alpha}}(\vx_{4})
\rangle \sim \frac{4}{\pi^4}\frac{{x_{13}}^{\mu}{x_{13}}^{\nu}}{{|\vx_{13}|}^{8}}
\frac{{x_{24}}^{\rho}\sigma^{\sigma}_{\alpha\dot{\alpha}}}{{|\vx_{24}|}^{4}}\langle
T_{\mu\nu}(\vx_{1})T_{\rho\sigma}(\vx_{2})\rangle \nonumber \\
=-\frac{16}{\pi^8}\frac{1}{{|\vx_{13}|}^{4}{|\vx_{24}|}^{4}{|\vx_{12}|}^{8}}
\left\{\frac{1}{2}\slashed{\vx_{24}}+\frac{2(\vx_{13}\cdot
J(\vx_{12})\cdot
\vx_{24})}{{|\vx_{12}|}^{2}}\frac{\slashed{\vx_{23}}\slashed{\tilde{x}_{31}}
\slashed{\vx_{14}}}{{|\vx_{13}|}^{2}}\right\}_{\alpha\dot{\alpha}}
\label{OPEcontse}
\end{eqnarray}
For the $u$-channel we use the $\mathcal{O}_{\tau}\bar{\Lambda}$ OPE
and its hermitian conjugate, and consider the limit in which
$\vx_{14}\rightarrow0$ and $\vx_{23}\rightarrow0$. The
short-distance expansion yields
\begin{eqnarray}
\langle
\mathcal{O}_{\tau}(\vx_{1})\Lambda_{\alpha}(\vx_{2})\mathcal{O}_{\bar{\tau}}(\vx_{3})
\bar{\Lambda}_{\dot{\alpha}}(\vx_{4})\rangle
\sim\frac{4}{\pi^4}\frac{{x_{14}}^{\mu}{x_{23}}^{\nu}}{{|\vx_{14}|}^{6}{|\vx_{23}|}^{6}}
\bar{\epsilon}\slashed{\tilde{x}_{14}}\langle
\Sigma_{\mu}(\vx_{1})\bar{\Sigma}_{\nu}(\vx_{2}) \rangle \nonumber
\slashed{\vx_{23}}
\\
=-\frac{4}{\pi^8}\frac{1}{{|\vx_{14}|}^{4}{|\vx_{23}|}^{4}{|\vx_{12}|}^{8}}
\left\{\frac{1}{2}\slashed{\vx_{24}}+\frac{2(\vx_{14}\cdot
J(\vx_{12})\cdot \vx_{23})}{{|\vx_{14}|}^{2}{|\vx_{23}|}^{2}}
\slashed{\vx_{23}}\slashed{\tilde{x}_{31}}\slashed{\vx_{14}}\right\}_{\alpha\dot{\alpha}}
\label{OPEcontsc}
\end{eqnarray}
So we can compare with the strongly coupled result, the operators in the gauge
theory side need to be properly normalized. Let us introduce the convenient normalization
\begin{equation}
\tilde{\mathcal{O}}_{\tau}(\vx)=\frac{\pi^2}{\sqrt{12}}\mathcal{O}_{\tau}(\vx)
\qquad
\tilde{\Lambda}(\vx)=\sqrt{\frac{2}{3}}\pi^2\Lambda(\vx)
\end{equation}
so the respective two-point functions have unitary coefficient.
\begin{equation}
\langle \tilde{\mathcal{O}}_{\tau}(\vx_{1})\tilde{\bar{\mathcal{O}}}_{\tau}(\vx_{2})\rangle=\frac{1}{|\vx_{12}|^8}
\qquad \qquad
\langle \tilde{\Lambda}(\vx_{1})\tilde{\bar{\Lambda}}(\vx_{2})\rangle=\frac{\slashed{\vx}_{12}}{|\vx_{12}|^8}
\end{equation}
This gives an overall factor of $\frac{\pi^8}{18}$ to the short-distance expansions (\ref{OPEcontse})
and (\ref{OPEcontsc}).

\subsection{Short--distance Expansion of the Supergravity Amplitude}
In this section we will determine the short--distance expansions for the
supergravity amplitude in terms of conformally invariant variables.
Using these we will be able to analyze the four-point function using
the double OPE structure that we uncovered in the previous section.
Furthermore, we will also compute the leading logarithmic
singularities, which signal the presence of semi-short operators
contributing to the OPE, and we will identify their anomalous
dimension. This is done explicitly in the $s$-channel, given that
the analysis is simpler and the identification is straightforward.

In general, a scalar quartic diagram can be decomposed in a regular
part and a singular part \cite{Dolan:2000ut}
\begin{equation}
\bar{D}_{\Delta_{1}\Delta_{2}\Delta_{3}\Delta_{4}}(u,v)=
\bar{D}_{\Delta_{1}\Delta_{2}\Delta_{3}\Delta_{4}}(u,v)_{reg}+
\bar{D}_{\Delta_{1}\Delta_{2}\Delta_{3}\Delta_{4}}(u,v)_{sing}
\end{equation}
where each term is given by series expansions in powers of $u$ and
$1-v$. The regular part will account for terms of the form $\ln u$
that lead to contributions to anomalous dimensions of order $1/N^2$.
Defining
\begin{equation}
s=\frac{1}{2}(\Delta_{1}+\Delta_{2}-\Delta_{3}-\Delta_{4})
\end{equation}
the explicit expressions for
$\bar{D}_{\Delta_{1}\Delta_{2}\Delta_{3}\Delta_{4}}(u,v)_{reg}$ and
$\bar{D}_{\Delta_{1}\Delta_{2}\Delta_{3}\Delta_{4}}(u,v)_{sing}$ are
\begin{eqnarray}
\bar{D}_{\Delta_{1}\Delta_{2}\Delta_{3}\Delta_{4}}(u,v)_{reg}&=&
\frac{(-1)^s}{s!}\frac{\Gamma(\Delta_{1})\Gamma(\Delta_{2})\Gamma(\Delta_{3}+s)\Gamma(\Delta_{4}+s)}{\Gamma(\Delta_{1}
+\Delta_{2})} \nonumber \\
&\times&\left( \sum_{m,n=0}^{\infty}\frac{(\Delta_{1})_{m}(\Delta_{4}+s)_{m}}{m!(s+1)_{m}}
\frac{(\Delta_{2})_{m+n}(\Delta_{3}+s)_{m+n}}{n!(\Delta_{1}+\Delta_{2})_{2m+n}}g_{mn}u^{m}(1-v)^{n} \right. \nonumber \\
&-& \Big. \ln{u}G(\Delta_{2},\Delta_{3}+s,1+s,\Delta_{1}+\Delta_{2};u,1-v) \Big)
\label{Dregular}
\end{eqnarray}
where
\begin{eqnarray}
g_{mn}&=&\psi(m+1)+\psi(s+m+1)+2\psi(\Delta_{1}+\Delta_{2}+2m+n)-\psi(\Delta_{1}+m)
\nonumber \\
&-&\psi(\Delta_{4}+s+m)-\psi(\Delta_{2}+m+n)-\psi(\Delta_{3}+s+m+n) \\
G(a,b,c,d,x,y)&=&\sum_{m,n=0}^{\infty}\frac{(d-a)_{m}(d-b)_{m}}{m!(c)_{m}}\frac{(a)_{m+n}(b)_{m+n}}{n!(d)_{2m+n}}x^{m}y^{n}
\end{eqnarray}
Here $\psi(x)$ is the digamma function. The singular part is given by
\begin{eqnarray}
\bar{D}_{\Delta_{1}\Delta_{2}\Delta_{3}\Delta_{4}}(u,v)_{sing}&=&u^{-s}
\frac{\Gamma(\Delta_{1}-s)\Gamma(\Delta_{2}-s)\Gamma(\Delta_{3})
\Gamma(\Delta_{4})}{\Gamma(\Delta_{3}+\Delta_{4})} \nonumber \\
&\times& \sum_{m=0}^{s-1}(-1)^{m}(s-m-1)!\frac{(\Delta_{1}-s)_{m}(\Delta_{2}-s)_{m}
(\Delta_{3})_{m}(\Delta_{4})_{m}}{m!(\Delta_{3}+\Delta_{4})_{2m}} \nonumber \\
&\times& u^{m}F(\Delta_{2}-s+m,\Delta_{3}+m, \Delta_{3}+\Delta_{4}+2m;1-v)
\label{Dsingular}
\end{eqnarray}
The cases in which $s<0$ can be taken into account by means of the identity
\begin{equation}
\Db(u,v)=u^{\Delta_{3}+\Delta_{4}-\Sigma}\bar{D}_{\Delta_{4}\Delta_{3}\Delta_{2}\Delta_{1}}(u,v)
\end{equation}
Convergence of the series is ensured given $u,1-v \sim 0$. Therefore
when analyzing a particular channel, one must ensure the conformal
ratios are defined so that (\ref{Dregular}) and (\ref{Dsingular})
behave appropriately, and this can be achieved by using the
different identities relating $\bar{D}$-functions, listed in the
appendix.

We now are ready to compute the asymptotic expansions for the amplitude, in the different channels. We adopt the
following coordinate choices
\begin{enumerate}
\item Graviton Channel (t-channel): $|\vx_{13}|^2 \to 0$ and $|\vx_{24}|^2 \to 0$,
which corresponds to $1/u \to 0$ and $v/u \to 1$.
\item Gravitino Channel (u-channel): $|\vx_{14}|^2 \to 0$ and $|\vx_{23}|^2 \to 0$,
which corresponds to $v \to 0$ and $u \to 1$.
\item Exotic Channel (s-channel): $|\vx_{12}|^2 \to 0$ and $|\vx_{34}|^2 \to 0$,
which corresponds to $u \to 0$ and $v \to 1$.
\end{enumerate}
We will now discuss each limit and their contributions to the singular and regular parts.

\subsubsection{Graviton Channel}
To analyze this channel one needs first to rewrite the amplitude
(\ref{finalamp}) in terms of the conformal coordinates
\begin{equation}
u'=\frac{1}{u}=\frac{|\vx_{13}|^{2}|\vx_{24}|^2}{|\vx_{12}|^2|\vx_{34}|^2} \qquad
v'=\frac{v}{u}=\frac{|\vx_{14}|^{2}|\vx_{23}|^2}{|\vx_{12}|^2|\vx_{34}|^2}
\end{equation}
To this effect, the identity that is required is
\begin{equation}
\bar{D}_{\Delta_{1}\Delta_{2}\Delta_{3}\Delta_{4}}(u,v)=u^{-\Delta_{2}}\bar{D}_{\Delta_{4}\Delta_{2}\Delta_{3}\Delta_{1}}(1/u,v/u)
\end{equation}
Using the formula (\ref{Dsingular}), one finds that the most
singular terms for the expansion are given by \footnote{We will drop the $\tilde{}$ on the understading
that all operators have two-point functions with unit coefficient.}
\begin{eqnarray}
&&\langle
\mathcal{O}_{\tau}(\vx_{1})\Lambda(\vx_{2})\mathcal{O}_{\bar{\tau}}(\vx_{3})\bar{\Lambda}(\vx_{4})\rangle_{t}\vert_{sing}
\nonumber \\
&&\sim - \frac{4}{5N^2}\frac{1}{|\vx_{13}|^{8}|\vx_{24}|^{8}}\left\{\left[\frac{u'}{9}{Y'}^2
-\frac{7{u'}^2}{36}+\cdots\right]\slashed{\vx}_{24}
+\left[-\frac{1}{6}{u'}^2
Y'+\cdots\right]\frac{\slashed{\vx}_{23}\slashed{\tilde{x}}_{31}\slashed{\vx}_{14}}{|\vx_{13}|^2}\right\}
\nonumber \\
&&=-\frac{4}{5N^2}\frac{1}{|\vx_{13}|^{4}|\vx_{24}|^{4}|\vx_{12}|^{8}}\left\{\left[\frac{1}{9}{u'}^{-1}{Y'}^2
+\cdots\right]\slashed{\vx}_{24}
+\left[-\frac{1}{6}
Y'+\cdots\right]\frac{\slashed{\vx}_{23}\slashed{\tilde{x}}_{31}\slashed{\vx}_{14}}{|\vx_{13}|^2}\right\}
\label{singtchannel}
\end{eqnarray}
where we introduced the variable $Y'\equiv 1-v'$. One can see here,
that the expected leading singularity coming from the stress-energy
tensor, is present given the dependence of the leading term on
${Y'}^2$. Now we can compare the singular terms in
(\ref{singtchannel}) with what is expected from the OPE
(\ref{OPEcontse}). It is convenient to define the following variable
\begin{equation}
t=\frac{|\vx_{12}|^{2}|\vx_{34}|^{2}-|\vx_{14}|^{2}|\vx_{23}|^{2}}{|\vx_{12}|^{2}|\vx_{34}|^{2}+|\vx_{14}|^{2}|\vx_{23}|^{2}}
= \frac{1-v'}{1+v'}
\label{tvar}
\end{equation}
Taking the $t$-channel limit, one can see that the leading term of $t$ is
\begin{equation}
t \sim -\frac{\vx_{13}\cdot J(\vx_{12}) \cdot \vx_{24}}{|\vx_{12}|^{2}}=\frac{1}{2}Y'+\cdots
\end{equation}
So now one can compare (\ref{OPEcontse}) and (\ref{singtchannel}).
Rewriting the free field theory result in terms of the variables
$u'$ and $Y'$ and re-introducing the dependence on $N$, one gets
\begin{eqnarray}
&&\langle
\mathcal{O}_{\tau}(\vx_{1})\Lambda(\vx_{2})\Obt(\vx_{3})\bar{\Lambda}(\vx_{4})\rangle
\nonumber \\
&&=-\frac{4}{N^2}\frac{1}{|\vx_{13}|^{4}|\vx_{24}|^{4}|\vx_{12}|^{8}}
\left\{\left[\frac{1}{9}+\cdots\right]\slashed{\vx}_{24}
+\left[-\frac{2}{9}
Y'+\cdots\right]\frac{\slashed{\vx}_{23}\slashed{\tilde{x}}_{31}\slashed{\vx}_{14}}{|\vx_{13}|^2}\right\}
\end{eqnarray}
One can see that the tensorial structure is very similar, and that
in the supergravity approximation, $T_{\mu\nu}^{free}= \frac{1}{5}T_{\mu\nu}$
\cite{Arutyunov:2000ku,Anselmi:1998ms}. This is expected
given that in the strongly couple regime, one only expects single
trace operators to give the leading order poles, whereas in the
free field theory result, long
operators are also present ($\mathcal{K}_{\mu\nu}$, which belongs to
the Konishi multiplet, and $\Xi_{\mu\nu}$ which is orthogonal to
both $T_{\mu\nu}$ and $\mathcal{K}_{\mu\nu}$. Both operators are
dual to string modes and decouple in the strong coupling limit).
This is easily seen from the fact that the free field result does not have
a ${Y'}^2$ term. \footnote{One could still be more careful, and consider the contributions to the free OPE
coming from the scalars and the fermions. In principle one should rewrite
the contributions to the stress-energy tensor in terms of an orthogonal basis,
and perform the computations keeping the long operators. In this way, one can
identify precisely which terms will dissapear when taking the large $N$ limit,
given that the long states decouple.}

Now we turn to the logarithmic terms. It is well-known that
supergravity tree-diagrams receive logarithmic corrections
\cite{Freedman:1998tz,Bianchi:1999ge}, which signal the presence of
composite operators arising from the approach of two chiral primary
operators. Moreover, one can read out the anomalous dimensions from
the coefficient of the logarithmic term. Schematically one has terms
in the asymptotic expansion of the form
\begin{equation}
u^{\frac{\Delta-l}{2}}Y^{l}\sim\frac{1}{2}\Delta^{(1)}u^{\frac{\Delta^{(0)}-l}{2}}Y^{l}\log{u}
\end{equation}
as $u\rightarrow0$, where $\Delta^{(0)}$ is the classical conformal
dimension and $\Delta^{(1)}$ is the quantum (anomalous) correction
to the conformal dimension. Using (\ref{Dregular}) to obtain the
regular part of the amplitude, one determines the logarithmic part
of the short--distance expansion
\begin{equation}
\langle
\mathcal{O}_{\tau}(\vx_{1})\Lambda(\vx_{2})\Obt(\vx_{3})\bar{\Lambda}(\vx_{4})\rangle_{t}\vert_{\log}\sim
-\frac{1}{N^2}\left\{\frac{10}{21}\slashed{\vx}_{24}
+\frac{64u'}{21}
\frac{\slashed{\vx}_{23}\slashed{\tilde{x}}_{31}\slashed{\vx}_{14}}{|\vx_{13}|^2}\right\}|\vx_{12}|^{4}\log u'+\cdots
\end{equation}
Given the absence of an appropriate conformal partial wave expansion
involving half-spin operators, is difficult to be precise on the
relation of these coefficients with the normalization constants to
the two- and three-point functions, and the anomalous dimensions of
the composite operators. However, it is possible to be more precise
in the case of the exotic channel that will be analyzed below.

\subsubsection{Gravitino Channel}
In this case, the amplitude (\ref{finalamp}) has to be reexpressed in terms of the ratios
\begin{equation}
u''=v=\frac{|\vx_{14}|^{2}|\vx_{23}|^2}{|\vx_{13}|^2|\vx_{24}|^2} \qquad
v''=u=\frac{|\vx_{12}|^{2}|\vx_{34}|^2}{|\vx_{13}|^2|\vx_{24}|^2}
\end{equation}
One needs then the identity
\begin{equation}
\bar{D}_{\Delta_{1}\Delta_{2}\Delta_{3}\Delta_{4}}(u,v)=\bar{D}_{\Delta_{3}\Delta_{2}\Delta_{1}\Delta_{4}}(v,u)
\end{equation}
The most singular terms are then
\begin{eqnarray}
&&\langle
\mathcal{O}_{\tau}(\vx_{1})\Lambda(\vx_{2})\Obt(\vx_{3})\bar{\Lambda}(\vx_{4})\rangle_{u}\vert_{sing}
\nonumber\\
&&\sim
\frac{1}{N^2}\frac{1}{|\vx_{13}|^{8}|\vx_{24}|^{8}}\left\{\left[-\frac{1}{9}\frac{1}{{u''}^2}
+\cdots\right]\slashed{\vx}_{24}
+\left[-\frac{2}{9}\frac{Y''}{{u''}^3}+\cdots
\right]\frac{\slashed{\vx}_{23}\slashed{\tilde{x}}_{31}\slashed{\vx}_{14}}{|\vx_{13}|^2}\right\}
\nonumber \\
&&=\frac{2}{9N^2}\frac{1}{|\vx_{14}|^{4}|\vx_{23}|^{4}|\vx_{12}|^{8}}\left\{\left[-\frac{1}{2}+\cdots\right]\slashed{\vx}_{24}
+
\left[-\frac{Y''}{u''}+\cdots\right]\frac{\slashed{\vx}_{23}\slashed{\tilde{x}}_{31}\slashed{\vx}_{14}}{|\vx_{13}|^2}\right\}
\label{singuchannel}
\end{eqnarray}
The leading order pole comes from the supercurrent. We analyze this
expression in the same way we did for the $t$-channel. Again it is
useful to use the analogous $t$ variable (\ref{tvar}) for this
channel, in which $\vx_{3}\leftrightarrow \vx_{4}$,
\begin{equation}
t'=\frac{|\vx_{12}|^{2}|\vx_{34}|^{2}-|\vx_{13}|^{2}|\vx_{24}|^{2}}{|\vx_{12}|^{2}|\vx_{34}|^{2}+|\vx_{13}|^{2}|\vx_{24}|^{2}}
= \frac{-Y''}{1+v''}
\end{equation}
so that when taking the $u$-channel limit, one gets
\begin{equation}
t'\sim -\frac{\vx_{14}\cdot J(\vx_{12}) \cdot
\vx_{23}}{|\vx_{12}|^{2}}=-\frac{1}{2}Y''+\cdots
\end{equation}
Rewriting what one gets from the free field theory OPE's, in terms
of $Y''$, one gets
\begin{eqnarray}
&&\langle\mathcal{O}_{\tau}(\vx_{1})\Lambda(\vx_{2})\Obt(\vx_{3})\bar{\Lambda}(\vx_{4})
\rangle \nonumber \\
&&=\frac{2}{9N^2}\frac{1}{{|\vx_{14}|}^{4}{|\vx_{23}|}^{4}{|\vx_{12}|}^{8}}\left\{\left[-\frac{1}{2}
+\cdots\right]\slashed{\vx_{24}}+\left[-\frac{
Y''}{u''}+\cdots\right]\frac{\slashed{\vx_{23}}\slashed{\tilde{x}_{31}}\slashed{\vx_{14}}}{|\vx_{13}|^2}\right\}
\end{eqnarray}
Comparing this expression to (\ref{singuchannel}) one sees that the
amplitude reproduces these terms.

The leading order logarithmic asymptotics are given below. This term
is to be interpreted as the $1/N^2$ renormalization effect to the
contribution from the composite operator $:\mathcal{O}_{\tau}\bar{\Lambda}:$
\begin{equation}
\langle
\mathcal{O}_{\tau}(\vx_{1})\Lambda(\vx_{2})\Obt(\vx_{3})\bar{\Lambda}(\vx_{4})\rangle_{u}\vert_{\log}\sim
-\frac{1}{N^2} \left\{\frac{10}{7}\slashed{\vx}_{24}
+\frac{20}{21}\frac{\slashed{\vx}_{23}\slashed{\tilde{x}}_{31}\slashed{\vx}_{14}}{|\vx_{13}|^2}\right\}
\frac{1}{|\vx_{12}|^{16}}\log u''+\cdots
\end{equation}
It would be interesting to give a precise interpretation to the
semi-short contributions to the $t$ and $u$-channel. We leave these
matters for future research.

\subsubsection{Exotic Channel}
The expansion is simplest to analyze in the exotic channel $\vx_{12}\to 0$,
$\vx_{34}\to 0$, since there are no
poles, so the leading order contribution to the short--distance expansion
is given by the logarithmic terms. The logarithmic terms are given
by
\begin{equation}
\langle
\mathcal{O}_{\tau}(\vx_{1})\Lambda(\vx_{2})\Obt(\vx_{3})\bar{\Lambda}(\vx_{4})\rangle_{s}\vert_{\log}\sim
\frac{1}{|\vx_{24}|^{16}}\slashed{\vx}_{24}\left\{-\frac{8}{N^2}\log
u+\cdots\right\}
\end{equation}
All terms are proportional to $\slashed{\vx}_{24}$ in this limit.
The structure of this term is precisely that of a double-trace
operator of dimension $15/2$. The structure of the term is such
that it is clearly expressed as a direct product of spin $1/2$ part and a scalar operator
(as in the bulk-to-boundary propagator for a spin $1/2$ operator, which is given by a
spin $1/2$ term times a scalar propagator).
The scalar part has conformal dimension 8, and
one can read off its anomalous dimension
immediately, given $\frac{1}{2}\Delta^{(1)}=-\frac{8}{N^2}$,
so that
\begin{equation}
\Delta^{(1)}=-\frac{16}{N^2}
\end{equation}
The scalar operator in this multiplet, which has $\Delta^{(0)}=4$,
is relevant and as noted in \cite{Arutyunov:2000ku},
can be used to study deformations of the $\mathcal{N}=4$ SYM.

As pointed out in \cite{D'Hoker:1999jp}, the space of operators of
approximate dimension 8 contains several semi-shorts distinguished
only by their $U(1)$ charge. In this case, there is mixing between
operators and we are only able to observe those that receive
corrections. However it is well-known that there is a particular
operator, which is known to be protected, and is a descendant of
that occurring in the tensor product of two chiral primaries in the
\textbf{20} of $SU(4)$ with $\Delta=4$ \cite{Arutyunov:2000ku}.
Shortening of semi-short operators is discussed in
\cite{Dolan:2002zh}, but as indicated there, there is no reason why
this operator has vanishing anomalous dimension.
\section{Discussion}
In this paper we derived the tree level four-point function of two dilatini and two
dilaton-axion fields in type IIB supergravity compactified in
$AdS_{5}\times S^{5}$, and we explicitly showed that its structure
is compatible with the double OPE expansion of $\mathcal{N}=4$ SYM
using AdS/CFT duality. Comparison of the asymptotic expansions of the supergravity
amplitude and the free-field theory results obtained from computing
the different OPE's, gave further evidence that long operators
decouple in the strong coupling limit, in a pattern reminiscent to
the one discussed in the cases of other four-point functions. Namely,
we were able to see that in the graviton channel, the free
field theory stress-energy tensor splits into orthogonal operators
belonging to different supersymmetry multiplets (hence the
difference in the normalization constant, and the apparent mismatch in
the coefficients of the expansion).
In the gravitino channel the results in the weakly coupled regime
and the strongly coupled one, suggest that there is no
splitting in the supercurrent. In general, one would expect the
free field theory supercurrent to split as
\begin{equation}
{\Sigma^{\mu}_{\alpha}}^{free}=c_{1}\Sigma^{\mu}_{\alpha}+c_{2}\Xi^{\mu}_{\alpha}
\end{equation}
where $\Xi^{\mu}_{\alpha}$ is orthogonal to $\Sigma^{\mu}_{\alpha}$
and $c_{1}+c_{2}=1$. This operator is dual to a string state
so it decouples from the theory in the supergravity approximation,
and does not appear in the strongly coupled YM OPE. However
the normalization constants of the free field operator and
the full YM supercurrent are the same, which is not consistent with the
fact that the split fields transform in different representations of
supersymmetry \cite{Arutyunov:2000ku}. This implies that either $c_{2}=O(1/N)$
or indeed that ${\Sigma^{\mu}_{\alpha}}^{free}=\Sigma^{\mu}_{\alpha}$.
It would be very interesting to investigate this issue in more detail\footnote{
Indeed, H. Osborn pointed out to us, after completion of this paper, that the supercurrent
is a unique operator and does not split when the theory becomes interacting, whereas
the stress-energy tensor is not unique, and one needs to go to an orthonormal basis
to do the analysis.}.

The analysis of the anomalous dimensions (and structure constants)
of the semi-short operators in cases involving fermionic operators
is not straigthforward, given that no conformal partial wave
expansion has been computed explicitly for half-spin operators.
The coefficients of the logarithmic terms arising in the graviton
and gravitino channels are then generically interpreted
as leading order $1/N^2$  corrections to the
anomalous dimensions and normalization constants of the two- and three-point
functions of double-trace operators. Precise determination of these
quantities is left as future work. However the fact that the exotic
channel contains no single-trace operators, and that the
single and triple gamma terms of the amplitude combine in this limit,
makes the analysis clearer, and in this
case it was possible to determine the anomalous dimension of the
semi-short operator. This operator is identified as a descendent of
$\mathcal{O}_{1}$ in \cite{Arutyunov:2000ku}. It is left as future
work, to see if this operator can be included in the discussion
of the integrated OPE truncation in \cite{Basu:2004dm}, given that
in the large $N$ limit, this operator is present and should be included
in the OPE.

Finally it is known from \cite{Drummond:2006by} that superconformal
symmetry should be enough to determine the precise form of the
amplitude from the knowledge of any other four-point function of $1/2$-BPS
operators in the current multiplet (either the four-point function
of CPO's or the four-point function of operators dual to the dilaton-axion
field). In a forthcoming paper \cite{Osborn:2007ho}, Osborn derives the result
presented in this paper from the knowledge of the correlator
$\langle \mathcal{O}_{\tau}\mathcal{O}_{\bar{\tau}}\mathcal{O}_{\tau}\mathcal{O}_{\bar{\tau}}\rangle$
computed in \cite{D'Hoker:1999pj}.
\appendix
\section{Properties of D-Functions}
In order to make this paper self-contained, we collect the general
properties and identities involving the $D$-functions.  These are
defined as integrals over $AdS_{5}$, by the formula
\begin{equation}
D_{\Delta_{1}\Delta_{2}\Delta_{3}\Delta_{4}}(\vx_{1},\vx_{2},\vx_{3},\vx_{4})=\int
\frac{d^{5}z}{z_{0}^{5}}\tilde{K}_{\Delta_{1}}(z,\vx_{1})\tilde{K}_{\Delta_{2}}(z,\vx_{2})
\tilde{K}_{\Delta_{3}}(z,\vx_{3})\tilde{K}_{\Delta_{4}}(z,\vx_{4})
\end{equation}
with
\begin{equation}
\tilde{K}_{\Delta}(z,\vx)=\left(\frac{z_{0}}{z_{0}^2+(\vec{z}-\vx)^{2}}\right)^{\Delta}
\end{equation}
D-integrals have also a representation in terms of integrals over
Feynman parameters
\begin{equation}
D_{\Delta_{1}\Delta_{2}\Delta_{3}\Delta_{4}}(\vx_{1},\vx_{2},\vx_{3},\vx_{4})
=\frac{\pi^{2}\Gamma(\Sigma-2)\Gamma(\Sigma)}{2\prod_{i}\Gamma(\Delta_{i})}\int
\prod_{j}d\alpha_{j}\alpha_{j}^{\Delta_{j}-1}\frac{\delta(\sum_{j}\alpha_{j}-1)}{(\sum_{k<l}\alpha_{k}
\alpha_{l}x_{kl}^2)^{\Sigma}}
\label{Dfeynmanp}
\end{equation}
where $2\Sigma=\sum_{i}\Delta_{i}$. Immediately one can see that any
$D$-function can be obtained by differentiation of the box-integral:
\begin{equation}
B(x_{ij})=\int
\prod_{j}d\alpha_{j}\frac{\delta(\sum_{j}\alpha_{j}-1)}{(\sum_{k<l}\alpha_{k}\alpha_{l}x_{kl}^2)^{\Sigma}}
\end{equation}
Using (\ref{Dfeynmanp}), one can derive the following identity
\begin{equation}
\frac{\p}{\p
x_{12}^2}D_{\Delta_{1}\Delta_{2}\Delta_{3}\Delta_{4}}=
-\frac{\Delta_{1}\Delta_{2}}{(\Sigma-2)}D_{\Delta_{1}+1\Delta_{2}+1\Delta_{3}\Delta_{4}}
\label{Didderiv}
\end{equation}
D-functions can be expressed in an inverted frame $\vx \to \vx'=
\vx/|\vx|^2$, in terms of $W$-functions. These are defined as
\begin{equation}
{W_{k}}^{\Delta'}(a,b)\equiv \int [dw]
\frac{w^{2\Delta'+2a+2k}_{0}}{w^{2k}}\frac{1}{(w-x)^{2\Delta'}}\frac{1}{(w-y)^{2\Delta'+2b}}
\end{equation}
A useful relation which was used in the text, was that of its
derivatives
\begin{eqnarray}
\p_{x_{i}}{W_{k}}^{\Delta'}(a,b)&=&-2x_{i}\frac{k\Delta'}{(k+\Delta'+a-2)}{W_{k+1}}^{\Delta'+1}(a-1,b-1)
\nonumber \\
&-&2(x-y)_{i}\frac{\Delta'(\Delta'+b)}{(k+\Delta'+a-2)}{W_{k}}^{\Delta'+1}(a,b)
\\
\p_{y_{i}}{W_{k}}^{\Delta'}(a,b)&=&-2y_{i}\frac{k(\Delta'+b)}{(k+\Delta'+a-2)}{W_{k+1}}^{\Delta'+1}(a,b+1)
\nonumber \\
&-&2(y-x)_{i}\frac{\Delta'(\Delta'+b)}{(k+\Delta'+a-2)}{W_{k}}^{\Delta'+1}(a,b)
\end{eqnarray}
We define now the $\bar{D}$-functions, which are functions of
conformal invariant ratios, $u$ and $v$, by
\begin{equation}
\bar{D}_{\Delta_{1}\Delta_{2}\Delta_{3}\Delta_{4}}(u,v)=\kappa
\frac{|\vx_{31}|^{2\Sigma-2\Delta_{4}}|\vx_{24}|^{2\Delta_{2}}}
{|\vx_{41}|^{2\Sigma-2\Delta_{1}-2\Delta_{4}}|\vx_{34}|^{2\Sigma-2\Delta_{3}-2\Delta_{4}}}
D_{\Delta_{1}\Delta_{2}\Delta_{3}\Delta_{4}}
\label{Dbardef}
\end{equation}
where
\begin{equation}
\kappa=\frac{2}{\pi^2}\frac{\Gamma(\Delta_{1})\Gamma(\Delta_{2})\Gamma(\Delta_{3})\Gamma(\Delta_{4})}{\Gamma(\Sigma-2)}
\end{equation}
One can obtain identities relating different $\bar{D}$-functions by
using the differentiation relation (\ref{Didderiv}). These are
\begin{eqnarray}
\bar{D}_{\Delta_{1}+1\Delta_{2}+1\Delta_{3}\Delta_{4}}&=&-\p_{u}\bar{D}_{\Delta_{1}\Delta_{2}\Delta_{3}\Delta_{4}}
\nonumber \\
\bar{D}_{\Delta_{1}\Delta_{2}+1\Delta_{3}+1\Delta_{4}}&=&-\p_{v}\bar{D}_{\Delta_{1}\Delta_{2}\Delta_{3}\Delta_{4}}
\nonumber \\
\bar{D}_{\Delta_{1}\Delta_{2}\Delta_{3}+1\Delta_{4}+1}&=&(\Delta_{3}
+\Delta_{4}-\Sigma-u\p_{u})\bar{D}_{\Delta_{1}\Delta_{2}\Delta_{3}\Delta_{4}}
\nonumber \\
\bar{D}_{\Delta_{1}+1\Delta_{2}\Delta_{3}\Delta_{4}+1}&=&(\Delta_{1}
+\Delta_{4}-\Sigma-v\p_{v})\bar{D}_{\Delta_{1}\Delta_{2}\Delta_{3}\Delta_{4}}
\nonumber \\
\bar{D}_{\Delta_{1}\Delta_{2}+1\Delta_{3}\Delta_{4}+1}&=&(\Delta_{2}
+u\p_{u}+v\p_{v})\bar{D}_{\Delta_{1}\Delta_{2}\Delta_{3}\Delta_{4}}
\nonumber \\
\bar{D}_{\Delta_{1}+1\Delta_{2}\Delta_{3}+1\Delta_{4}}&=&(\Sigma-\Delta_{4}
+u\p_{u}+v\p_{v})\bar{D}_{\Delta_{1}\Delta_{2}\Delta_{3}\Delta_{4}}
\label{Dids1}
\end{eqnarray}
There are additional identities which relate $\bar{D}$-functions
with different values of $\Sigma$, and can be derived by repeated
use of $(\ref{Dids1})$. These are
\begin{eqnarray}
(\Delta_{2}+\Delta_{4}-\Sigma)\bar{D}_{\Delta_{1}\Delta_{2}\Delta_{3}\Delta_{4}}
&=&\bar{D}_{\Delta_{1}\Delta_{2}+1\Delta_{3}\Delta_{4}+1}
-\bar{D}_{\Delta_{1}+1\Delta_{2}\Delta_{3}+1\Delta_{4}} \nonumber \\
(\Delta_{1}+\Delta_{4}-\Sigma)\bar{D}_{\Delta_{1}\Delta_{2}\Delta_{3}\Delta_{4}}
&=&\bar{D}_{\Delta_{1}+1\Delta_{2}\Delta_{3}\Delta_{4}+1}
-v\bar{D}_{\Delta_{1}\Delta_{2}+1\Delta_{3}+1\Delta_{4}} \nonumber \\
(\Delta_{3}+\Delta_{4}-\Sigma)\bar{D}_{\Delta_{1}\Delta_{2}\Delta_{3}\Delta_{4}}
&=&\bar{D}_{\Delta_{1}\Delta_{2}\Delta_{3}+1\Delta_{4}+1}
-u\bar{D}_{\Delta_{1}+1\Delta_{2}+1\Delta_{3}\Delta_{4}}
\end{eqnarray}
Furthermore, there are identities relating $\bar{D}$-functions with
the same $\Sigma$. The most frequently used is
\begin{equation}
\Delta_{4}\bar{D}_{\Delta_{1}\Delta_{2}\Delta_{3}\Delta_{4}}=\bar{D}_{\Delta_{1}\Delta_{2}\Delta_{3}+1\Delta_{4}+1}
+\bar{D}_{\Delta_{1}\Delta_{2}+1\Delta_{3}\Delta_{4}+1}
+\bar{D}_{\Delta_{1}+1\Delta_{2}\Delta_{3}\Delta_{4}+1}
\end{equation}
More non-trivial identities can be derived by using the previous
ones. In simplifying the graviton exchange we made use of the
identities
\begin{eqnarray}
\bar{D}_{\Delta+2\Delta_{1}\Delta\Delta_{2}}+\bar{D}_{\Delta\Delta_{1}\Delta+2\Delta_{2}}
&=&(\Delta_{2}-1)(\Delta_{2}-2)\bar{D}_{\Delta\Delta_{1}\Delta\Delta_{2}-2}
-2(\Delta_{2}-1)\bar{D}_{\Delta\Delta_{1}+1\Delta\Delta_{2}-1}
\nonumber \\
&&+\bar{D}_{\Delta\Delta_{1}+2\Delta\Delta_{2}}-2\bar{D}_{\Delta+1\Delta_{1}\Delta+1\Delta_{2}}
\label{nontrivialid1}
\end{eqnarray}
\begin{eqnarray}
2\bar{D}_{\Delta_{1}+1\Delta_{2}\Delta_{1}+1\Delta_{2}+2}
&=&\left(\frac{\Gamma(\Delta_{1}+1)}{\Gamma(\Delta_{1})}\right)^{2}\bar{D}_{\Delta_{1}\Delta_{2}\Delta_{1}\Delta_{2}}
-(u+v)\bar{D}_{\Delta_{1}+1\Delta_{2}+1\Delta_{1}+1\Delta_{2}+1}
\nonumber \\
&&-\frac{(2\Delta_{1}+1)}{\Gamma(\Delta_{1})^{2}}\bar{D}_{\Delta_{1}+1\Delta_{2}\Delta_{1}+1\Delta_{2}}
+\bar{D}_{\Delta_{1}+2\Delta_{2}\Delta_{1}+2\Delta_{2}}
\label{nontrivialid2}
\end{eqnarray}
Finally, we comment on the various symmetries that these functions
exhibit. By means of conformal symmetry, one can see that
\begin{eqnarray}
\bar{D}_{\Delta_{1}\Delta_{2}\Delta_{3}\Delta_{4}}(u,v)
&=&v^{-\Delta_{2}}\bar{D}_{\Delta_{1}\Delta_{2}\Delta_{4}\Delta_{3}}(u/v,1/v)
\nonumber \\
\bar{D}_{\Delta_{1}\Delta_{2}\Delta_{3}\Delta_{4}}(u,v)
&=&v^{\Delta_{4}-\Sigma}\bar{D}_{\Delta_{2}\Delta_{1}\Delta_{3}\Delta_{4}}(u/v,1/v)
\nonumber \\
\bar{D}_{\Delta_{1}\Delta_{2}\Delta_{3}\Delta_{4}}(u,v)
&=&v^{\Delta_{1}+\Delta_{4}-\Sigma}\bar{D}_{\Delta_{2}\Delta_{1}\Delta_{4}\Delta_{3}}(u,v)
\nonumber \\
\bar{D}_{\Delta_{1}\Delta_{2}\Delta_{3}\Delta_{4}}(u,v)
&=&u^{\Delta_{3}+\Delta_{4}-\Sigma}\bar{D}_{\Delta_{4}\Delta_{3}\Delta_{2}\Delta_{1}}(u,v)
\nonumber \\
\bar{D}_{\Delta_{1}\Delta_{2}\Delta_{3}\Delta_{4}}(u,v)
&=&\bar{D}_{\Delta_{3}\Delta_{2}\Delta_{1}\Delta_{4}}(v,u)
\nonumber \\
\bar{D}_{\Delta_{1}\Delta_{2}\Delta_{3}\Delta_{4}}(u,v)
&=&\bar{D}_{\Sigma-\Delta_{3}\Sigma-\Delta_{4}\Sigma-\Delta_{1}\Sigma-\Delta_{2}}(u,v)
\end{eqnarray}

\section{Some Manipulations involving $\bar{D}$-functions}
Here we show how the direct results from the supergravity
computation can be simplified using the identities (\ref{Dids1}),
(\ref{nontrivialid1}) and (\ref{nontrivialid2}) given in the previous appendix.

We start by working out the triple gamma matrix contribution to the
graviton amplitude (\ref{Gravitondirect}). We start by rewriting the
first four terms, using the identities
\begin{eqnarray}
\bar{D}_{3425}&=&\frac{1}{2}(\bar{D}_{3526}-\bar{D}_{4435}) \qquad \qquad
\bar{D}_{4435}=\frac{1}{2}(\bar{D}_{2536}-\bar{D}_{3445})
\nonumber \\
\bar{D}_{4435}&=&\bar{D}_{4536}-\bar{D}_{5445} \hspace{21.mm}
\bar{D}_{4354}=\bar{D}_{3546}-\bar{D}_{4455}
\end{eqnarray}
Collecting terms, one can see that the third identity above for
$\bar{D}_{4435}$ and $\bar{D}_{3445}=\bar{D}_{3546}-\bar{D}_{4455}$
can be used again. This yields the final expression in (\ref{amp1}).
The same manipulation is also employed in the terms contained in the
single gamma matrix part.

Less trivial identities are required to simplify the additional
terms on the single gamma matrix contribution. One starts by
using (\ref{nontrivialid2}) on the last three $\bar{D}$-functions, so that
\begin{eqnarray}
2\bar{D}_{2426}&=&\bar{D}_{1414}-3\bar{D}_{2424}+\bar{D}_{3434}-(u+v)\bar{D}_{2525} \nonumber \\
2\bar{D}_{5254}&=&4\bar{D}_{2424}-5\bar{D}_{3434}+\bar{D}_{4444}-(u+v)\bar{D}_{3535} \nonumber \\
2\bar{D}_{2426}&=&9\bar{D}_{3434}-7\bar{D}_{4444}+\bar{D}_{5454}-(u+v)\bar{D}_{4545} \nonumber
\end{eqnarray}
and (\ref{nontrivialid1}) on the pairs
\begin{eqnarray}
\bar{D}_{3416}+\bar{D}_{4163}&=&2\bar{D}_{1414}-4\bar{D}_{2424}+\bar{D}_{3434}-2\bar{D}_{2426} \nonumber \\
\bar{D}_{4426}+\bar{D}_{4264}&=&6\bar{D}_{2424}-6\bar{D}_{3434}+\bar{D}_{4444}-2\bar{D}_{5454} \nonumber \\
\bar{D}_{5436}+\bar{D}_{4365}&=&12\bar{D}_{3434}-8\bar{D}_{4444}+\bar{D}_{4545}-2\bar{D}_{5355} \nonumber
\end{eqnarray}
Substitution of these expressions in (\ref{Gravitondirect}) yields
the final result for the graviton exchange diagram (\ref{amp1}).

Now we turn to the gravitino amplitude. To simplify the single
gamma matrix contribution, one needs to use
the following identities
\begin{eqnarray}
\bar{D}_{3452}+\bar{D}_{2453}&=&2\bar{D}_{2442}-\bar{D}_{3443} \nonumber \\
\bar{D}_{4462}+\bar{D}_{3463}&=&3\bar{D}_{2543}-\bar{D}_{3544} \nonumber \\
\bar{D}_{4453}+\bar{D}_{3454}&=&3\bar{D}_{3443}-\bar{D}_{4444} \nonumber \\
\bar{D}_{5463}+\bar{D}_{4464}&=&4\bar{D}_{3544}-\bar{D}_{5454} \nonumber
\end{eqnarray}
Direct substitution on (\ref{Gravitinodirect}) give the single gamma
contribution specified on (\ref{amp2}). Simplification of the triple
gamma matrix term is even simpler. One just needs to use the
identities
\begin{eqnarray}
\bar{D}_{3461}&=&\bar{D}_{2451}+\bar{D}_{2552} \nonumber \\
\bar{D}_{4462}&=&\bar{D}_{3452}+\bar{D}_{3553} \nonumber \\
\bar{D}_{5463}&=&\bar{D}_{4453}+\bar{D}_{4554} \nonumber
\end{eqnarray}
to obtain what is given in the final result.

\acknowledgments I am very grateful to Prof. Michael Green for
numerous discussions and encouragement on the completion of this
project. I would also like to thank Anirban Basu for collaboration
at an initial stage of this work, and to Ling-Yang Hung and
Rui F. Lima Matos for discussions. Finally, I would like to thank
Prof. Hugh Osborn for discussing his results prior to publication and
for his careful reading of the manuscript and his valuable comments
on the contents of this paper.

This work has been supported by CONACyT Mexico and the ORSAS UK scheme.

\bibliographystyle{JHEP}
\bibliography{bibfile}

\end{document}